\documentclass[useAMS,usenatbib]{mnras_2020}
\usepackage{graphicx,amsmath,url,natbib}

\usepackage{multirow}
\usepackage[flushleft]{threeparttable}
\usepackage{xcolor}
\usepackage{ulem}

\newcommand{\gtrsim}{\,\raisebox{-0.4ex}{$\stackrel{>}{\scriptstyle\sim}$}\,}
\newcommand{\lesssim}{\,\raisebox{-0.4ex}{$\stackrel{<}{\scriptstyle\sim}$}\,}
\newcommand{\ergs}{\mbox{ erg s}^{-1}}

\newcommand{\up}{\mbox{up}}
\newcommand{\dn}{\mbox{dn}}
\newcommand{\maxrm}{\rm max}

\newcommand{\minrm}{\rm min}
\newcommand{\eq}[1]{Eq.~(\ref{#1})}

\newcommand{\gammamax}{\gamma_{\rm max}}

\begin{document}

\title[Evolution of non-thermal electrons]{Simulating the dynamics and synchrotron emission from relativistic jets II. Evolution of non-thermal electrons}
\author[Mukherjee et al.]{Dipanjan Mukherjee$^1$\thanks{dipanjan@iucaa.in}, Gianluigi Bodo$^3$, Paola Rossi$^3$, Andrea Mignone$^2$ \& Bhargav Vaidya $^4$ \\
$^{1}$ Inter-University Centre for Astronomy and Astrophysics, Post Bag 4, Pune - 411007, India. \\
$^{2}$ Dipartimento di Fisica Generale, Universita degli Studi di Torino , Via Pietro Giuria 1, 10125 Torino, Italy\\
$^{3}$ INAF/Osservatorio Astrofisico di Torino, Strada Osservatorio 20, I-10025 Pino Torinese, Italy\\
$^{4}$ Discipline of Astronomy, Astrophysics and Space Engineering, Indian Institute of Technology Indore, \\
       Khandwa Road, Simrol, 453552, India
}
\date{\today}
\pagerange{\pageref{firstpage}--\pageref{lastpage}} 
\pubyear{2015}
\maketitle

\begin{abstract}
We have simulated the evolution of non-thermal cosmic ray electrons (CREs) in 3D relativistic magneto hydrodynamic (MHD) jets evolved up to a height of 9 kpc. The CREs have been evolved in space and in energy concurrently with the relativistic jet fluid, duly accounting for radiative losses and acceleration at shocks. We show that jets stable to MHD instabilities show expected trends of regular flow of CREs in the jet spine and acceleration at a hotspot followed by a settling backflow. However, unstable jets create complex shock structures at the jet-head (kink instability), the jet spine-cocoon interface and the cocoon itself (Kelvin-Helmholtz modes). CREs after exiting jet-head  undergo further shock crossings in such scenarios and are re-accelerated in the cocoon. CREs with different trajectories in turbulent cocoons have different evolutionary history with different spectral parameters. Thus at the same spatial location, there is mixing of different CRE populations, resulting in a complex total CRE spectrum when averaged over a given area. Cocoons of unstable jets can have an excess build up of energetic electrons due to re-acceleration at turbulence driven shocks and slowed expansion of the decelerated jet. This will add to the non-thermal energy budget of the cocoon.
\end{abstract}


\section{Introduction}\label{sec:intro}
Supermassive blackholes at the centres of galaxies can launch powerful relativistic jets that can grow to extragalactic scales. These structures have since long been observed in radio frequencies, starting from observations of Cygnus A by \citet{jennison53a}. Since early days it had been established that incoherent synchrotron emission by energetic non-thermal electrons in magnetised jets gives rise to the observed radiation  \citep{shklovskii55a,burbidge56a}. Subsequently, there has been extensive work to understand the physical nature of these objects \citep[e.g.][]{blandford74a,blandford79a} and the source of the emission. We refer the reader to reviews such as \citet{begelman84a}, \citet{worrall06a} and \citet{blandford19a} for more elaborate discussions on the historical evolution of the concept of relativistic radio jets and their emission processes.

A key factor that strongly affects the emitted radiation is the nature and evolution of the non-thermal electrons inside the jet that gives rise to the synchrotron emission. It had again been identified quite early that the  electrons must be energised inside the jet after ejection from the central nucleus, to avoid losing energy due to adiabatic losses in the expanding jet \citep{longair73a,blandford79a}. The electrons are primarily accelerated at strong shocks inside the jet and its cocoon via diffusive shock acceleration \citep[or Fermi 1st order processes,][]{blandford78a,drury83a,blandford87a}. However, recent works \citep[e.g.][]{rieger07a,sironi21a} have also pointed to the importance of other microphysical processes such as Fermi second order and reconnection at shear layers to accelerate particles to some extent. After an acceleration event, the energetic electrons are expected to suffer cooling due to synchrotron and inverse-Compton losses. 

Detailed semi-analytic calculations of the evolution of the energy spectrum of non-thermal electron populations moving in a fluid flow have been presented in several papers \citep{kardashev62a,jaffe73a,murgia99a,hardcastle13a}. However, there remain several restrictive assumptions in such semi-analytic approaches regarding the nature of the magnetic field distribution experienced by the particles, the pitch angle of the electron with respect to the magnetic field and the location of acceleration. The non-thermal electrons can experience a wide variety of physical parameters of the background environment during their trajectory, which shapes their spectrum. Addressing this requires a fully numerical approach where the non-thermal electron population are allowed to sample the local properties of the fluid, as opposed to averaged quantities. 

In some recent works, numerical simulations have been presented that self-consistently evolve spectra of non-thermal electrons concurrently with the fluid \citep[viz.][]{jones99a,micono99a,mimica09a,fromm16a,vaidya18a,winner19a,huber21a}. The standard approach involves solving the phase space evolution of the non-thermal electrons while accounting for radiative losses and acceleration due to microphysical processes such as diffusive shock acceleration. However, such methods  have been applied to study astrophysical jets in radio galaxies by only a handful of papers, and with restrictive assumptions. For example, \citet{jones99a,micono99a,tregillis01a} have simulated the evolution of non-thermal electrons in large scale jets but in a non-relativistic framework. Other works have evolved the non-thermal electrons embedded in relativistic fluid \citep[e.g.][]{mimica09a,fromm16a}, but for small parsec scale quasi-steady state jets, without exploring the detailed evolution up to larger distances (kpc), which is relevant for jets in radio galaxies.

In this series of papers, we will study the properties of the non-thermal electrons, evolved concurrently with the relativistic jet fluid, while duly accounting for the radiative energy losses and acceleration at shocks \citep{vaidya18a}. The current paper, second in the series, presents the behaviour and evolution of the non-thermal electrons along with the bulk relativistic jet flow for different simulations with a wide range of jet parameters. The results on the dynamics of these simulated jets have been presented earlier in \citep{mukherjee20a}. In subsequent papers we shall discuss the impact on observations (synchrotron and inverse-Compton) expected from such simulated jets.

The plan of the paper is as follows. In Sec.~\ref{sec.CREintro} we summarise the implementation of the method to solve for the evolution of non-thermal electrons presented in \citet{vaidya18a} and new changes introduced in this paper in the modelling of diffusive shock acceleration. In Sec.~\ref{sec.inject} we discuss the numerical details of injection of non-thermal electrons. In Sec.~\ref{sec.paperI} we briefly summarise the results of the dynamics of the simulated jets, which have been presented in detail in \citet{mukherjee20a}. In Sec.~\ref{sec.results} we discuss the results of the spatial and spectral evolution of the non-thermal electrons and how they vary for different simulations with different jet parameters. In Sec.~\ref{sec.discuss} we summarise our results and discuss on their implications.

\section{Evolution of the Cosmic Ray Electrons}\label{sec.CREintro}
In \citet[][hereafter BV18]{vaidya18a} a detailed analytical and numerical framework was developed to evolve  a distribution of non-thermal particles both spatially and in momentum space. This was introduced as the \textsc{Lagrangian Particle} module in the \textsc{PLUTO} code \citep{mignone07}. This module has been applied in the current work to study the evolution of non-thermal electrons in relativistic jets. The non-thermal relativistic electrons at a region in space are modelled as 
ensembles of macro-particles which we refer to as a cosmic ray electron macro-particle or CRE in short. The CRE are advected spatially following the fluid motions. Their energy is found by evolving their phase space distribution function, duly accounting for energy losses due to radiative processes such as synchrotron emission and inverse-Compton interaction with CMB photons. Appendix~\ref{append.bv18} briefly summarises the method of phase space evolution of a CRE.

\subsection{Diffusive shock acceleration}\label{sec.dsa} 
Electrons crossing shocks can be accelerated to higher energies via diffusive shock acceleration  \citep[hereafter DSA, see ][for a review]{blandford78a,drury83a}. The DSA theory predicts that particles exiting a shock can have a power-law spectrum whose index depends on the shock compression ratio defined in \eq{eq.cmpr}. The maximum energy of the resultant spectrum is found by equating the time scale of Synchrotron driven radiative losses to the acceleration time scale (as given in \eq{eq.gammamax}). For relativistic shocks, power-law index of the spectrum is calculated differently for parallel\footnote{A parallel shock has its normal close to the down-stream magnetic field. The shock normal and downstream magnetic field are at large angles for perpendicular shocks. } and perpendicular shocks, as in \eq{eq.qpar} and \eq{eq.qperp}. Further details of the above steps are summarised in Appendix~\ref{append.DSA}. In the next sections we describe an empirical approach to obtain the downstream spectrum of a shocked CRE.

\subsubsection{Previous approaches of evaluating shocked spectra}
In several previous similar works \citep[e.g.][]{mimica09a,bottcher10a,fromm16a,vaidya18a}, the spectrum of shock accelerated particles were reset to a power-law with an index predicted by the theory of DSA. This would result in the particles losing the history of their spectra before the entering the shock. The normalisation of the spectra and the lower-limit ($E_{\minrm}$) were set by enforcing the particles to have a fractional energy  and number density of the fluid. However, this has the following two disadvantages. 
\begin{itemize}
\item In complex simulations with multiple shock features, a single computational cell may host more than one particle, and only a fraction of them may have crossed a shock. Some CRE macro-particles may have been advected to the cell without crossing a shock, by different flow stream lines. Additionally there can be multiple particles that exit the same shock feature at the same time, thus requiring a simultaneous spectral update. Hence, in a departure from \citet{vaidya18a}, where each particle was assigned an energy of $f_E \times \rho \epsilon$, we distribute to the newly shocked particles in the cell only the energy remaining after subtracting from $f_E\times \rho \epsilon$  the existing cosmic ray energy density .

\item Secondly, resetting the spectrum to a new power-law results in total loss of the spectral history of previous shock encounters. Such multiple shock crossings can leave an imprint on the final spectra due to different acceleration time scales and power-law indices generated at different shocks \citep[e.g.][]{melrose93b,micono99a,giesler2000a,meli13a,parker14a}. This is especially problematic in a scenario where a particle having crossed a strong shock (higher shock compression ratio $r$), subsequently passes through a weaker shock with a steeper power-law index as per DSA theory. Replacing the earlier shallower spectrum with a new steeper power-law,  will result in an artificial loss of energy of the CRE macro-particle.

\end{itemize}

To overcome the above limitations we introduce a empirical approach to update the spectrum of a shocked CRE, as outlined in the next section.

\subsubsection{A empirical convolution based approach for updating CRE spectra}\label{sec.convol}
A particle enters a shock with an up-stream (pre-shock) spectrum $N_{\up}(E)$. On exiting the shock, the distribution is updated to a down-stream (post-shock) spectrum $N_{\dn}(E)$ as:
\begin{align}
N_{\dn}(E) &= \mathcal{C} \int_{\rm E_{min}}^E N_{\up}(E') G_{\rm DSA}(E,E')  \frac{dE'}{E'} \nonumber \\ 
           &= \mathcal{C} \int_{\rm E_{min}}^E N_{\up}(E') \left(\frac{E}{E'}\right)^{-q+2} \frac{dE'}{E'} \label{eq.dsa}
\end{align}
In \eq{eq.dsa} above, $N_{\up}(E)$ is convolved with a function $G_{\rm DSA}(E,E') = (E/E')^{-q+2}$, for $E \in (E_{\minrm}, E_{\maxrm})$. The minimum energy $E_{\minrm}$ is kept fixed to the minimum energy of the up-stream spectrum. The maximum energy $E_{\maxrm}$ is computed from \eq{eq.gammamax} and \eq{eq.acc} and the spectral index $q$  from \eq{eq.qpar} and \eq{eq.qperp}.

For non-relativistic shocks, \eq{eq.dsa} with $\mathcal{C} = q$ and $q = 3r/(r-1)$, is an exact representation of the down-stream spectrum \citep{blandford78a,drury83a}, obtained by solving for the phase space distribution function of the particles across the shock front. In the present work we extend this as a phenomenological approach to update the up-stream spectrum of  relativistic shocks. The methodology of \eq{eq.dsa} is similar to the implementation of re-accelerated particles in \citet{micono99a} and \citet{winner19a}. Approximate analytical solutions of the particle distribution at a relativistic shock without an incident source, predicts a power-law spectrum for the downstream particles \citep{takamoto15a}. Convolving the up-stream spectrum with such a power-law distribution is thus an empirical approach that accounts for the spectrum of the incident distribution, as a CRE exits a shock. Appendix~\ref{append.dsa} has further details regarding the choice of the power-law index predicted from the DSA theory.  

The above scheme is also similar in nature to the implementation of shock acceleration in \citet{jones99a} and \citet{tregillis01a}, where the spectrum in an energy bin has been updated only if the index of the local piece-wise power-law is lower than that predicted by DSA. This duly avoids artificial steepening of spectra when energised particles with flatter spectrum crosses a weak shock, which is also naturally taken care of by the convolution process used here, as also elaborated later in this section. 

The normalisation constant $\mathcal{C}$ is set in two steps, which ensure that the maximum local energy and number density of the CREs are less than or equal to a specified fraction of the fluid.  
\begin{itemize}
\item \textbf{Step 1: }  At computational cells where CREs are to be updated on exiting the shock, we enforce the  criterion that the total energy density of all CREs in the cell should be a fraction $f_E$ of the fluid internal energy density\footnote{It is to be noted that as a particle exits a shock into the downstream region, there may exist other particles inside the computational cell which may have been advected there without passing through a shock, or are remnants of a previous shock-exit.}. We thus first compute the total available energy in a computational volume that may be distributed amongst the newly shocked particles:
\begin{equation}
\Delta E = f_E \times(\rho \epsilon) - \sum_i \varepsilon_{i} \label{eq.fracE},
\end{equation}
 Here ($\rho \epsilon$) is the fluid internal energy density, given by
\begin{align}
\rho \epsilon &= p/(\Gamma - 1) \quad \mbox{ (Ideal EOS)} \\
\rho \epsilon &= (3/2)p -\rho c^2 + \sqrt{(9/4)p^2 + \rho^2 c^4} \quad \mbox{ (T-M EOS)} \label{eq.TM},
\end{align}
where \eq{eq.TM} is valid for the Taub-Mathews (T-M) equation of state \citep{taub48a,mignone07a}. 

The energy density of a single CRE macro-particle is given by $\varepsilon_i = \int E N(E) dE$.  Freshly shocked particles that are to be updated are excluded from the summation in the second term of \eq{eq.fracE}. For $\Delta E > 0$, the difference in energy is equally distributed to all shocked particles in the computational cell, which sets the value of the normalisation $\mathcal C$ in \eq{eq.dsa}. For $\Delta E \leq 0$, no new particles are updated inside the given cell, as the CRE energy density of existing particles have already reached the threshold of $f_E\times\left(\rho\epsilon\right)$.

\item \textbf{Step2: }
Subsequently, we compute 
\begin{equation}
\Delta N = f_N \times \rho/(\mu m_a) - \sum_i N^T_i \label{eq.fracN},
\end{equation}
 which is the difference between the fluid number density multiplied by a mass fraction $f_N$ and the total number density of all CREs in the cell, including that of shocked particles calculated in step 1 above. Here $m_a$ is the atomic mass unit and $\mu=0.6$ is the mean molecular weight for ionised gas. The number density of a CRE particle is obtained from $N^T_i = \int N(E) dE$. For $\Delta N < 0$, the normalisation $\mathcal{C}$ is reduced to  ensure that the total cosmic ray electron number density in a computational cell is $N_e = f_N \times \rho/(\mu m_a)$. 
\end{itemize}

To summarise, the normalisation is first set to preserve the energy densities of the particles to be a fraction of ($f_E$) the fluid energy density. Additionally, if the chosen normalisation results in the CRE number density becoming higher than a fraction ($f_N$) of the fluid number density, the previously obtained normalisation is lowered to ensure an exact match. In that case, the total CRE energy density will be lower than the  threshold of $f_E\times\rho\epsilon$, due to a lower value of the new normalisation. Thus the above two-step procedure ensures that the CREs have energies and number densities that are within the prescribed fractions $f_E$ and $f_N$ for energy and mass respectively.  For this work we have assumed $f_E = f_N = 0.1$. 

We must note here that chosen fractions are somewhat arbitrary, although  not far from the indications obtained from PIC simulations  \citep[see e.g.][]{sironi13a,caprioli14a,guo14a,Marcowith16a,marcowith20a}. In addition, though small, they are not negligible. In reality, CREs accelerated at shocks will subtract the corresponding fractional gain in energy from the fluid internal energy,  which  is not considered here. Feedback from radiative losses due to non-thermal particles can  change the shock structure, as demonstrated in \citet{bromberg09a} and \citet{bodo18a}. Such effects are not considered in this work as the CRE particles are considered to be passively evolving with the fluid. However, given that the CRE energy and mass fractions are restricted to a maximum of 10\%, the effects are not expected to be substantial, and the qualitative trends are not expected to be affected. Understanding the quantitative impact of energy losses and cosmic ray pressure requires detailed modelling of the interaction of the CRE with the fluid  and will be deferred to a future work.

\begin{figure}
	\centering
	\includegraphics[width = 7 cm, keepaspectratio] 
	{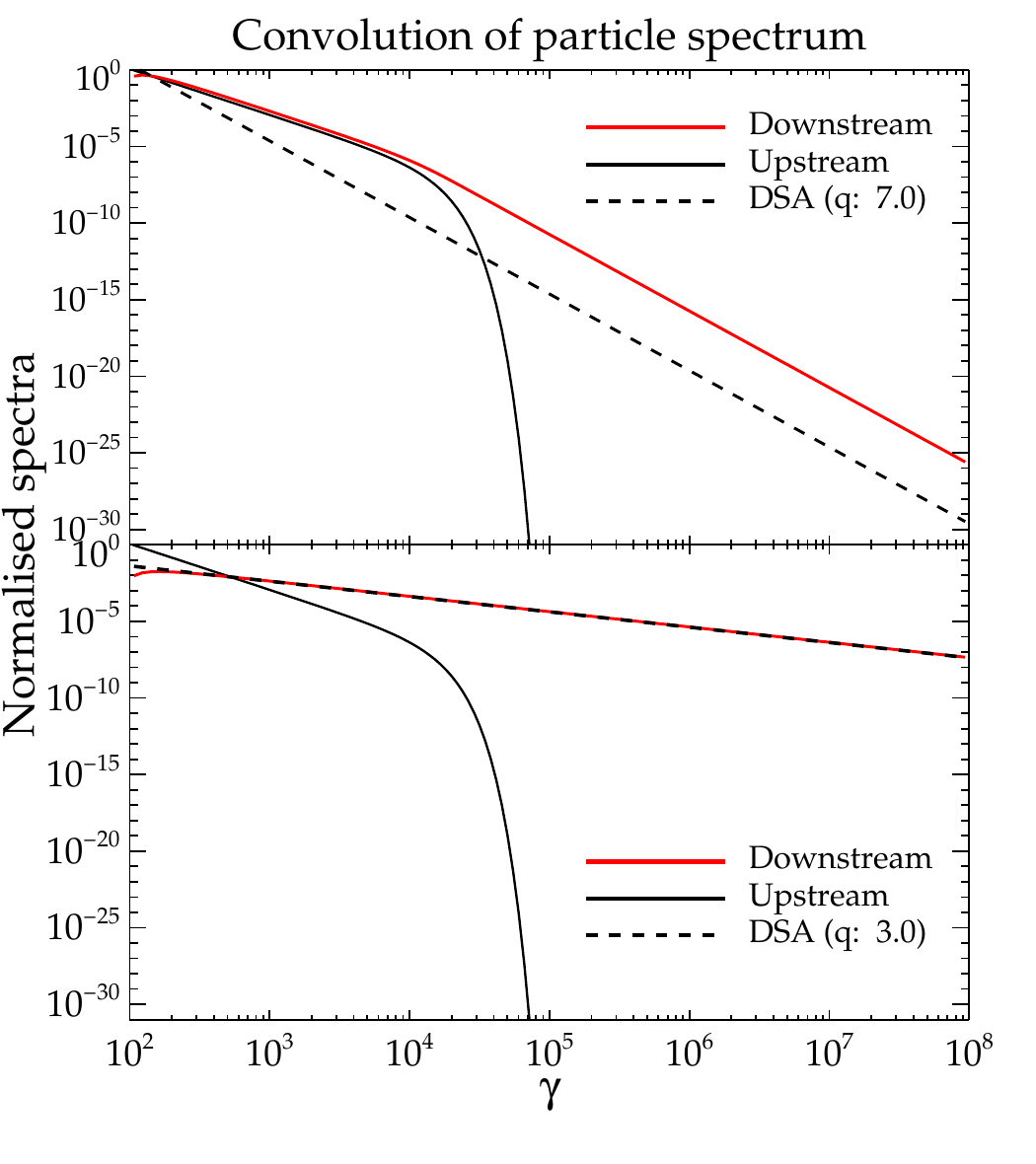}
        \caption{\small A demonstration of the empirical scheme involving convolution of the particle spectrum with the DSA predicted power-law function (in dashed). The up-stream spectrum (in black) of the particle before the shock update is a power-law with an exponential cut-off given by \eq{eq.mockdsa}. The resultant down-stream spectrum (in red) is obtained from their convolution. }
	\label{fig.mockdsa}
\end{figure}
The above procedure to update the spectrum of a shocked CRE is better at accounting for situations where a CRE has traversed multiple shocks of different strengths. The convolution procedure introduced here \citep[as also in][]{micono99a,winner19a} will redistribute lower energy electrons to higher energy bins as per the functional form of the spectrum predicted by DSA. This may be interpreted as the lower energy electrons being accelerated to higher energies. Thus in the previously described situation of a particle passing first through a stronger shock followed by a weaker one, the resultant spectrum would be a broken power-law.

The top panel of Fig.~\ref{fig.mockdsa} demonstrates two possible scenarios of shock crossing by a CRE. Here the up-stream spectrum is given by a power-law with an exponential cut-off:
\begin{equation}
N \propto \gamma_e^{-\alpha} \exp\left(-\frac{\gamma_e}{\gamma_c}\right) \label{eq.mockdsa},
\end{equation}
with $\alpha = 5$ and $\gamma_c = 10^4$ representing the slope at lower energies and cut-off energy respectively. The resultant convolution with a spectrum of $N \propto \gamma_e^{-7}$,  gives a broken power-law, as shown in the upper panel of Fig.~\ref{fig.mockdsa}. The lower energies below the break remain a power-law with $N \propto \gamma_e^{-5}$, whereas the higher energies with an initial exponential decay, gets populated with a power-law spectrum of $N \propto \gamma_e^{-7}$. Similarly, for a convolution with a power-law function of $N \propto \gamma_e^{-3}$,  the resultant spectrum yields a power-law of $N \propto \gamma_e^{-3}$ for the whole range.

\section{Numerical implementation of CRE injection}\label{sec.inject}
\begin{figure}
	\centering
	\includegraphics[height = 8 cm, keepaspectratio] 
	{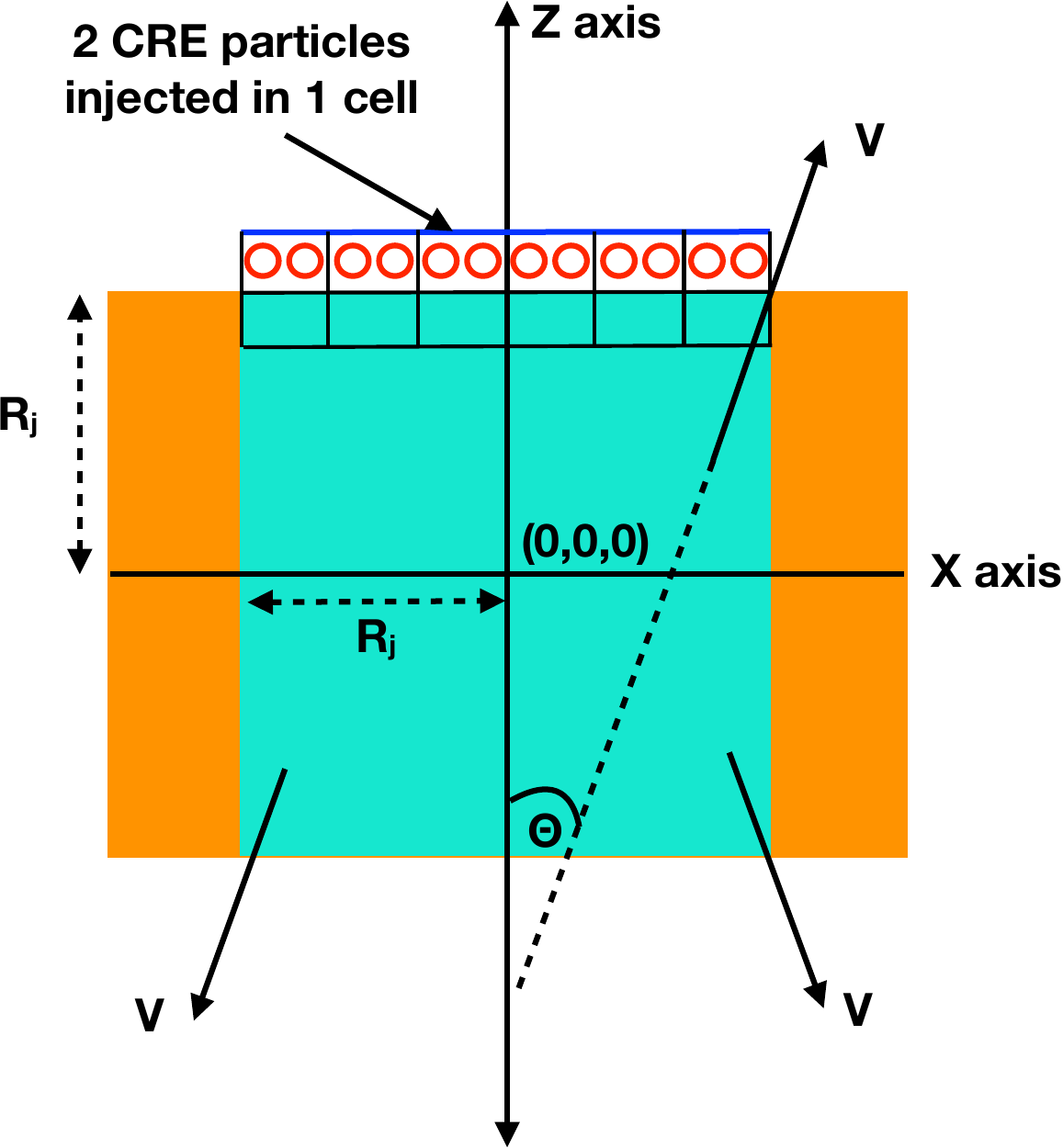}
        \caption{\small A representation of the injection zone of the jet in the simulation domain \citep[similar to Fig. 2 of ][]{mukherjee20a}, which also shows the injection of the cosmic ray electron (CRE) macro-particles. Two CREs, represented in red, are injected at every time step in a computational cell (denoted by the black boxes) just above the jet inlet, along the $Z$ axis. The area of injection spans up to the extent of the jet radius in the $X-Y$ plane. For one sided jets, CREs are injected only along the upper surface of the injection zone, as demarcated here. For twin jets (simulation E), CREs are injected along both upper and lower surfaces. }
	\label{fig.jetcartoon}
\end{figure}
We have performed a series of simulations of relativistic jets expanding into an ambient medium in \citet[][ hereafter Paper I]{mukherjee20a}. In this section we summarise the details of how CRE macro-particles are introduced into the simulation domain. We refer the reader to Sec.~2 of Paper I for details regarding the magneto-hydrodynamics of the simulation and related information regarding the simulation set up.

The CRE are injected in the simulation in computational zones just above the inlet of the jet along the $Z$ axis, as shown in Fig.~\ref{fig.jetcartoon}. Two CRE particles were injected at each cell, at each time step, to achieve greater filling of the cocoon volume. The injected particles are immediately advected by the jet flow upwards along the jet axis. Over the course of the simulations, typically several million particles are injected, as shown by the last column in Table~\ref{tab.sims}. 

At injection the CRE macro-particles are initialised with a steep power-law spectrum  ($\alpha_1 = 9$) and limiting Lorentz factors: $(\gamma_{\minrm},\gamma_{\maxrm}) \equiv (10^2,10^6)$ as
\begin{align}
&N(E) = n_m \left(\frac{1-\alpha_1}{E_{\maxrm}^{1-\alpha_1} - E_{\minrm}^{1-\alpha_1}}\right) E^{-\alpha_1} \label{eq.injectspec} \\
&\mbox{ where } \int^{E_{\maxrm}}_{E_{\minrm}} N(E) dE = n_m
\end{align}
Here $n_m$ is the CRE number density, whose value is set as follows. With $n_{pi}$ CRE particles injected in a computational cell with volume $\Delta V$, the total energy of the injected CREs is
\begin{align}
E_{i} &= n_{pi}\Delta V \int^{E_{\maxrm}}_{E_{\minrm}} E N(E) dE \\
 &\simeq  n_{pi}n_m m_e c^2 \Delta V \left(\frac{\alpha_1-1}{\alpha_1-2}\right) \gamma_{\minrm}  = f_E \left(\frac{\Gamma }{\Gamma -1}\right) p_j \Delta V\label{eq.en},
\end{align}
where $\gamma_{\maxrm} \gg \gamma_{\minrm}$ has been assumed. The last equality in \eq{eq.en} arises from assuming that the CRE energy density in the volume $\Delta V$ accounts for a fraction $f_E$ of the fluid enthalpy in rest frame. The jet pressure is given by $p_j$. Simplifying \eq{eq.en}, we get the CRE number density as
\begin{equation}
n_m = f_E \left(\frac{\Gamma}{\Gamma -1}\right) \frac{(\alpha_1-2)}{(\alpha_1-1)} \frac{p_j}{n_{pi} \gamma_{\minrm} m_e c^2} \label{eq.nm}.
\end{equation}

However, one must note that the starting configuration of the CRE spectrum defined in \eq{eq.injectspec} and \eq{eq.nm} is renewed when they encounter the first shock, which occurs fairly close to the injection surface, from the recollimation shocks within the jets. Thus the resultant spectra of the particles inside the main jet flow after some distance away from the injection surface are insensitive to the initial values.

\section{Review of results from Paper I}\label{sec.paperI}
\begin{table}
\caption{List of simulations and parameters}\label{tab.sims}

\centering
\begin{tabular}{| l | c | c | c | c |}
\hline
Sim. 	       &$\gamma_j $  & $\sigma_B$  & $P_j$ 		     &  No. Particles$^a$  \\
label	       &             &             & ($10^{45}$ergs$^{-1}$)  &  ($\times 10^6$)    \\ 
\hline
B$^b$          & 3           & 0.1         & $0.2$                  &   $16.4$    \\
D              & 5           & 0.01        & $1.1$                  &   $6.5$  \\
E              & 5           & 0.05        & $1.2$                  &   $18.0$  \\
F              & 5           & 0.1         & $1.2$                  &   $3.2$ \\
G              & 6           & 0.2         & $8.3$                  &   $14.9$    \\
H              & 10          & 0.2         & $16.4$                  &   $9.8$ \\
J              & 10          & 0.1         & $15.1$                  &   $55.3$  \\
\hline
\end{tabular} 

\begin{tablenotes}
\item $^a$ \small{The total number of particles injected during the course of the simulation.}
\item $^b$ \small{The density contrast defined as the ratio of the jet density to ambient gas is $4\times10^{-5}$. For all other cases, the value is $\eta=10^{-4}$ }
\end{tablenotes}

\flushleft

Parameters: \\
\begin{tabular}{|l | l|}
    $\gamma_b$:          & Jet Lorentz factor. \\
    $\sigma_B$:          & Jet magnetisation, the ratio of jet Poynting flux \\
                         & to  enthalpy flux. See eq.~(7) of Paper I. \\
    $P_j$:               & Jet power. See eq.~(9) of Paper I. \\
        
\end{tabular}
\end{table}
In \citet[][ Paper I]{mukherjee20a} we have presented the results of the simulations of relativistic jets, focusing on the evolution of the dynamics and fluid parameters in the jet and its cocoon. Table~\ref{tab.sims} lists the simulations where the non-thermal CREs were evolved along with the fluid. We keep the same nomenclature of the simulations as done in Paper I, for consistency. The simulations probe a wide range of jet parameters, namely: jet power ($P_j$), jet magnetisation ($\sigma_j$) and the bulk Lorentz factor ($\gamma_j$). Broadly we can divide the results into three groups: simulation B with a low power jet ($P_j \sim 10^{44} \ergs$), simulations D, E, F with moderate power jets ($P_j \sim 10^{45} \ergs$) and simulations G, H, J with high power jets ($P_j \sim 10^{46}\ergs$). In most simulations the jets were followed up to 9 kpc, except for simulation J where the domain is larger. 

In Paper I, we have discussed the dynamics of the jets. One of the primary results of the paper, which is also strongly relevant for the discussion in the current work, is the onset of different MHD instabilities for different ranges of jet parameters. We summarise briefly below the main findings of Paper I which are relevant to the present work. 

\begin{enumerate}
\item \textbf{Kink unstable (simulation B):} Simulation B with a low power jet ($P_j\sim10^{44} \ergs$) and strong magnetisation ($\sigma_B \sim 0.1$) is unstable to kink mode instabilities. The kink instabilities cause the jet axis and head to bend strongly, which creates complex shock structures at the jet head (see Fig.~4 in Paper I).  

\item \textbf{KH unstable (simulation D):} Jets with lower magnetisation, such as simulation D with $P_j \sim 10^{45} \ergs$ and $\sigma_B = 0.01$, suffer from Kelvin-Helmholtz (hereafter KH) instabilities. This decelerates the jet and creates turbulence inside the jet and cocoon, with a disrupted jet-spine due to vortical fluid motions resulting from the KH modes. 

\item \textbf{Stable jets (simulations F,H):} Moderate to high power jets ($P_j\sim 10^{45}-10^{46}\ergs$) with stronger magnetisation ($\sigma_B \sim 0.1$) remain stable to both kink and KH modes up to the run time of the simulations explored in these works, such as simulations F and H. This is due to reduced growth rates of KH modes due to stronger magnetic field and also reduced growth of both kink and KH modes due to higher bulk Lorentz factors.

\item \textbf{Turbulent cocoon (simulation G):} Higher pressure of the jet results in  more internal structure. The higher sound speed facilitates the growth of small scale perturbations \citep{rosen99a}, which can develop internal shocks and turbulence. Simulation G ($P_j \sim 10^{46} \ergs$) with an injected jet pressure five times that of simulation H, shows such internal instabilities and turbulence.
\end{enumerate}

In the rest of the paper, a major theme would be to explore the effect of the above different MHD instabilities on the evolution of the CREs.

\section{Results}\label{sec.results}
\begin{figure}
	\centering
	\includegraphics[width = 6. cm, keepaspectratio]{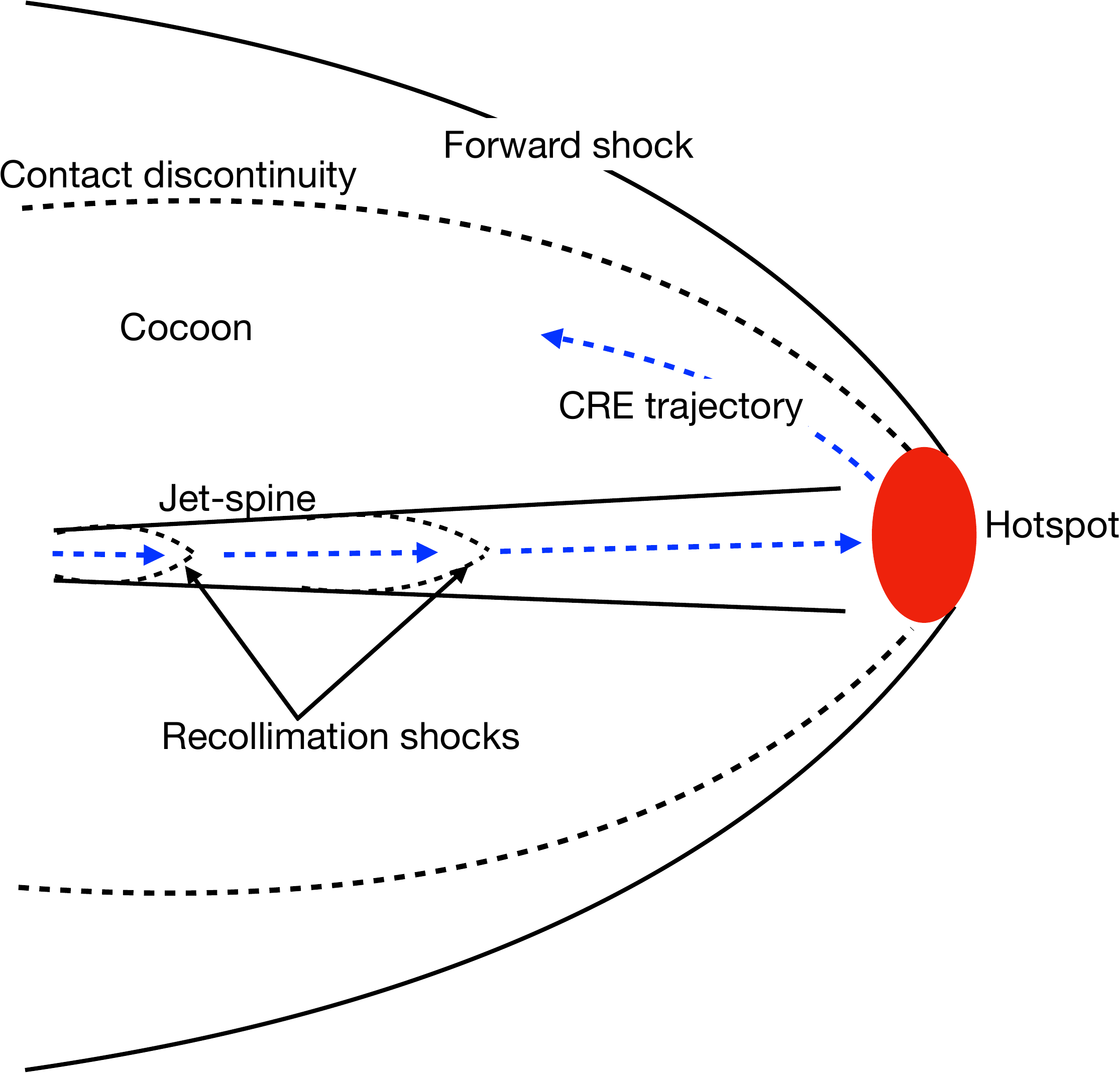}
        \caption{\small A schematic showing the structure of the jet spine, the cocoon surrounding it, the contact discontinuity and the forward shock. The non-thermal electrons (called here as cosmic ray electrons or CREs) are expected to travel first through the jet spine up to the hotspot and flow back down into the cocoon. The CREs are first accelerated at the sites of recollimation shocks \citep{norman82a,komissarov98a} inside the jet before they reach the strong shock at the hotspot.}
	\label{fig.jetCartoon}
\end{figure}
According to the standard evolutionary picture, CREs are considered to be carried by the jet up to the termination shock at the jet-head, and eventually follow the back-flow in the cocoon to lower heights, as shown in Fig.~\ref{fig.jetCartoon}. The particles are expected to be accelerated to high energies with a shallow power-law spectrum at the jet-head and subsequently cool down in the backflow inside the cocoon due to radiative losses from synchrotron and inverse-Compton interaction with the CMB radiation \citep{jaffe73a,murgia99a,harwood13a,harwood15a,hardcastle13a}. However, the above picture is simplistic, especially for complex flow  patterns arising out of MHD instabilities in the jet and the cocoon. In the following sub-sections, we describe the behaviour of the particles as they evolve with the jet and highlight the differences from the above standard paradigm.

\subsection{Evolution of CREs in the jet and cocoon}
\subsubsection{CREs in stable powerful jets}
\begin{figure}
	\centering
	\includegraphics[width = 6. cm, keepaspectratio] 
	{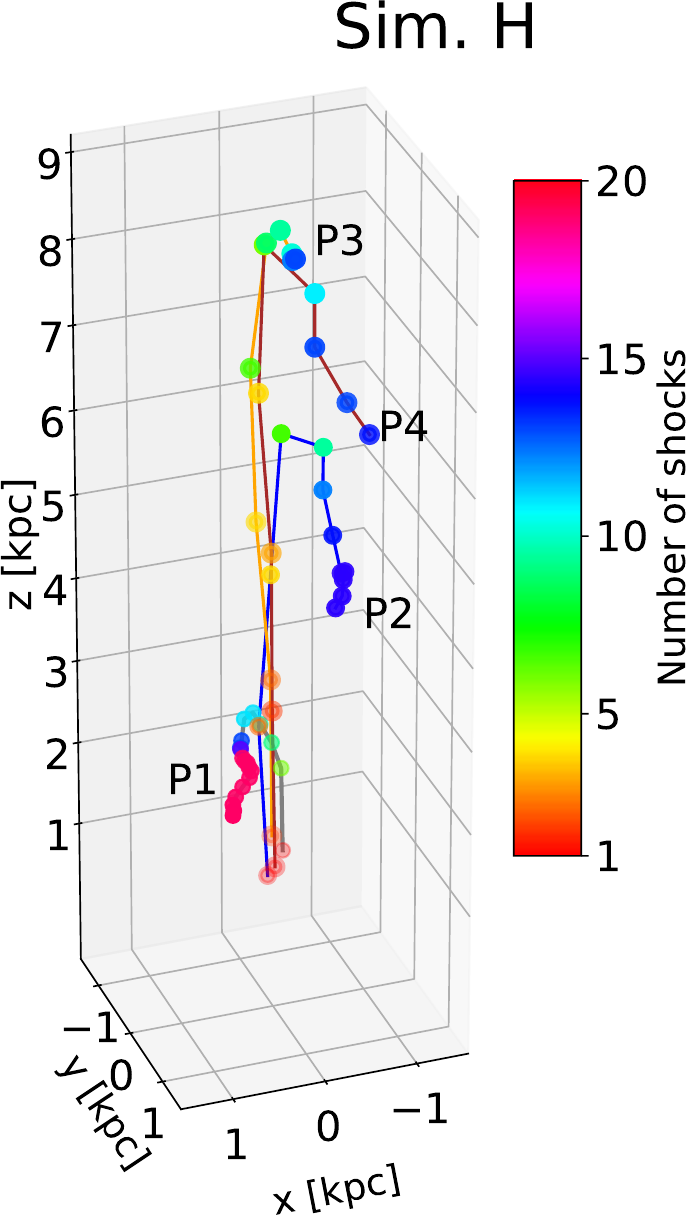}
        \caption{\small Trajectories of 4 representative particles (named P1, P2, P3 and P4) that undergo multiple shocks in simulation H. The color of the particles show the number of shocks crossed at a given point of the track.}
	\label{fig.phistSimH}
\end{figure}
\begin{figure}
	\centering
	\includegraphics[height = 6.2 cm, keepaspectratio] 
	{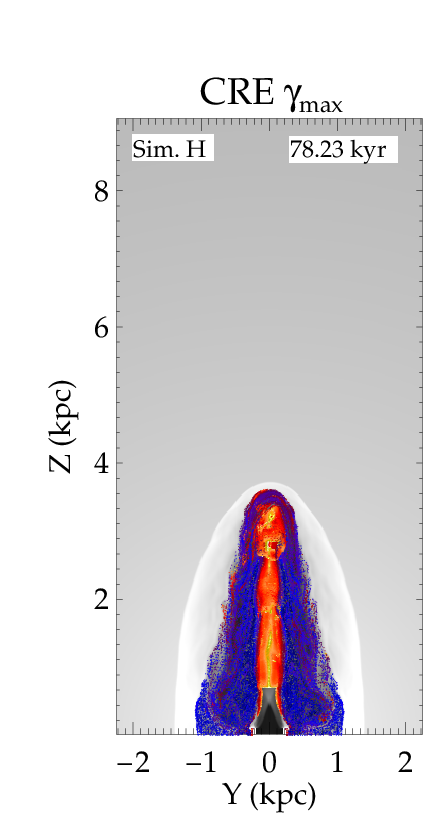}
	\includegraphics[height = 6.2 cm, keepaspectratio] 
	{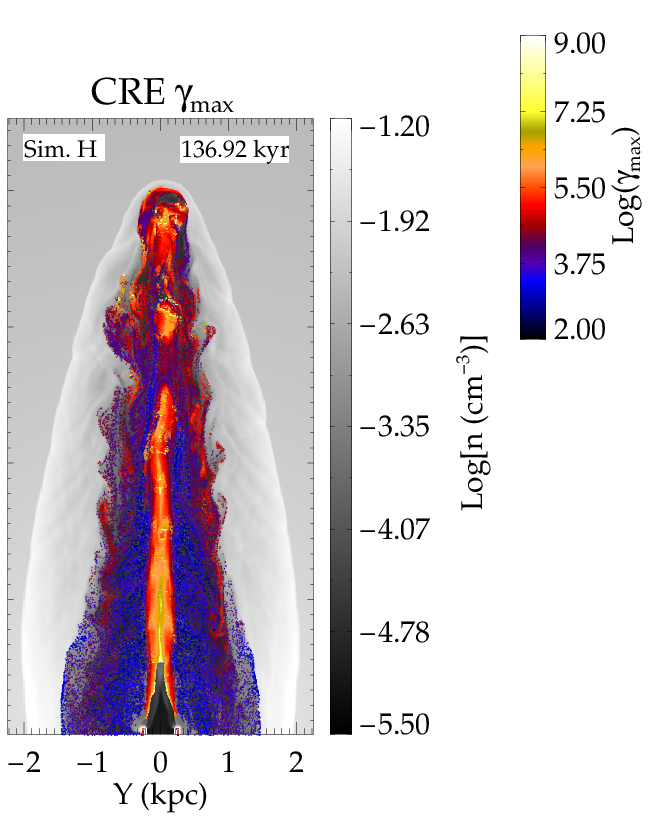}
	\includegraphics[height = 6.2 cm, keepaspectratio] 
	{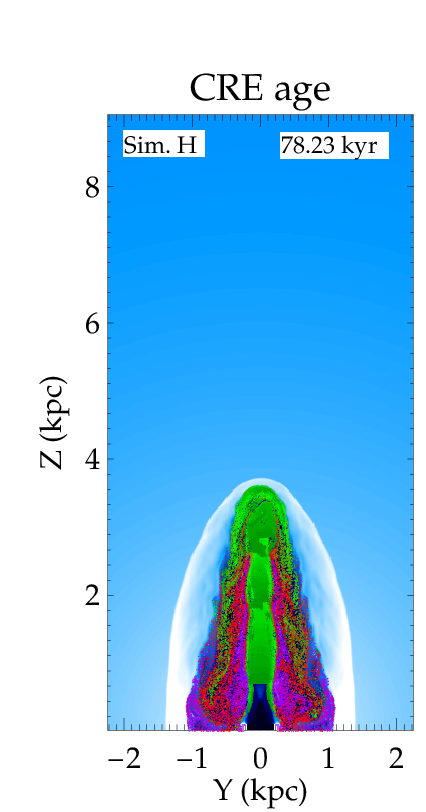}
	\includegraphics[height = 6.2 cm, keepaspectratio] 
	{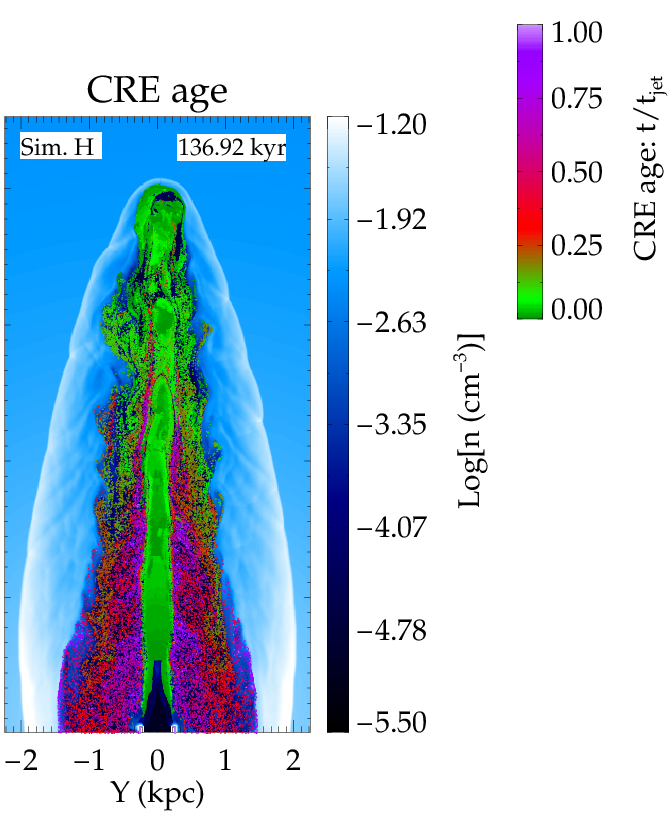}
        \caption{\small \textbf{Top panels:} Density slices in the $X-Z$ plane for simulation H at two different times. Over-plotted on the density maps are particles with colors representing the $\log(\gamma_{\maxrm})$, where $\gamma_{\maxrm}$ is the highest Lorentz factor of the CRE spectrum  \textbf{Bottom panels:} Density slices in the $X-Z$ plane with the CRE positions over-plotted. The color of the CRE represent the time since last shock, normalised to the time of the simulation. Freshly shocked particles (greenish hue) are near the jet head, with older particles (purple/pink) at the base.  }
	\label{fig.page2D}
\end{figure}
In this section we describe the coevolution of the CREs with the jet for case H, a prototypical stable jet with $P_{\rm jet} = 10^{46} \ergs$. The CREs injected at each time step (as outlined in Sec.~\ref{sec.inject}) are advected along with the jet flow till they reach the jet head. There, the CREs interact with the termination shock and turn back into the cocoon pushed by the jet's back-flow. This is well demonstrated by the trajectories of four representative CREs in Fig.~\ref{fig.phistSimH}.  The chosen particles from four different heights at the end of the simulation have the largest number of shock crossings  at the given heights.

The CREs are first accelerated inside the jet spine itself, at the several recollimation shocks, before they reach the jet head. This causes the particles to attain a large value of maximum Lorentz factor $\gamma_{\maxrm}$ inside the jet axis. This can be seen in the top panel of Fig.~\ref{fig.page2D}, where we present the CREs with color scaled to the value of $\log(\gamma_{\maxrm})$, over-plotted on the density profile in the $Y-Z$ plane. The bottom panels of Fig.~\ref{fig.page2D} show the particles coloured by time lapsed since last shock encounter. The CREs within the jet axis are also seen to be young in age as they have been relatively recently shocked at the sites of recollimation.

Once the particles reach the jet-head, they cross the strong shock at the Mach-disk and follow the backflow of the jet into the cocoon. The cocoon has older particles, represented in magenta and red in Fig.~\ref{fig.page2D}. The CREs in the cocoon also have lower maximum energies as they lose energy steadily due to radiative losses. Thus the CREs in this simulation of a jet stable to MHD instabilities conform well to the standard evolutionary picture proposed in earlier works \citep[e.g.][]{jaffe73a,komissarov98a,turner15a}, as also summarised in the schematic in Fig.~\ref{fig.jetCartoon}.

\subsubsection{Impact of MHD instabilities}\label{sec.instablities}
\begin{itemize}
\item \emph{Simulation B, kink unstable:}
\begin{figure*}
	\centering
	\includegraphics[width = 8. cm, keepaspectratio] 
	{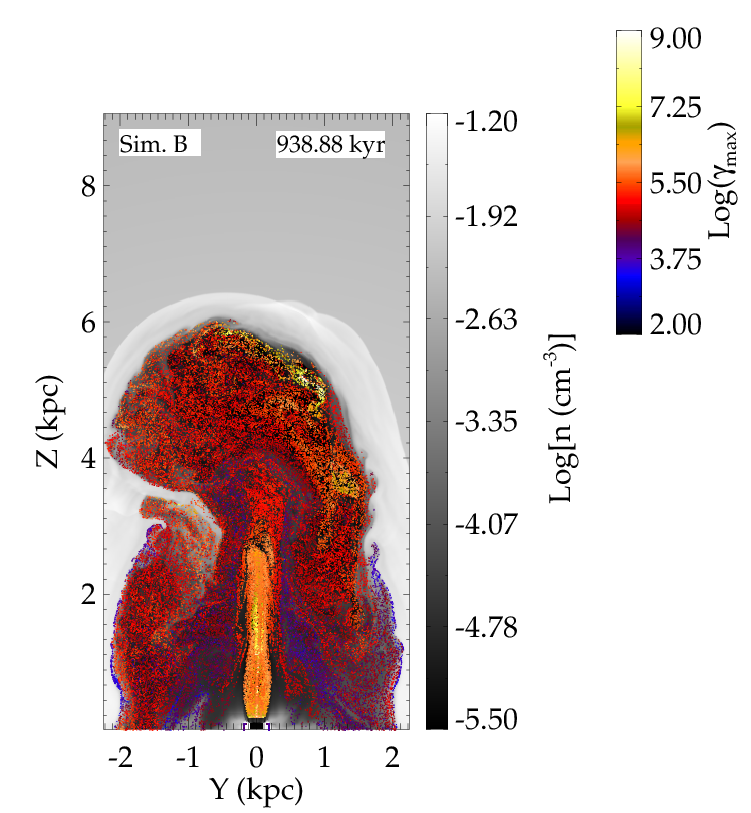}
	\includegraphics[width = 7. cm, keepaspectratio] 
	{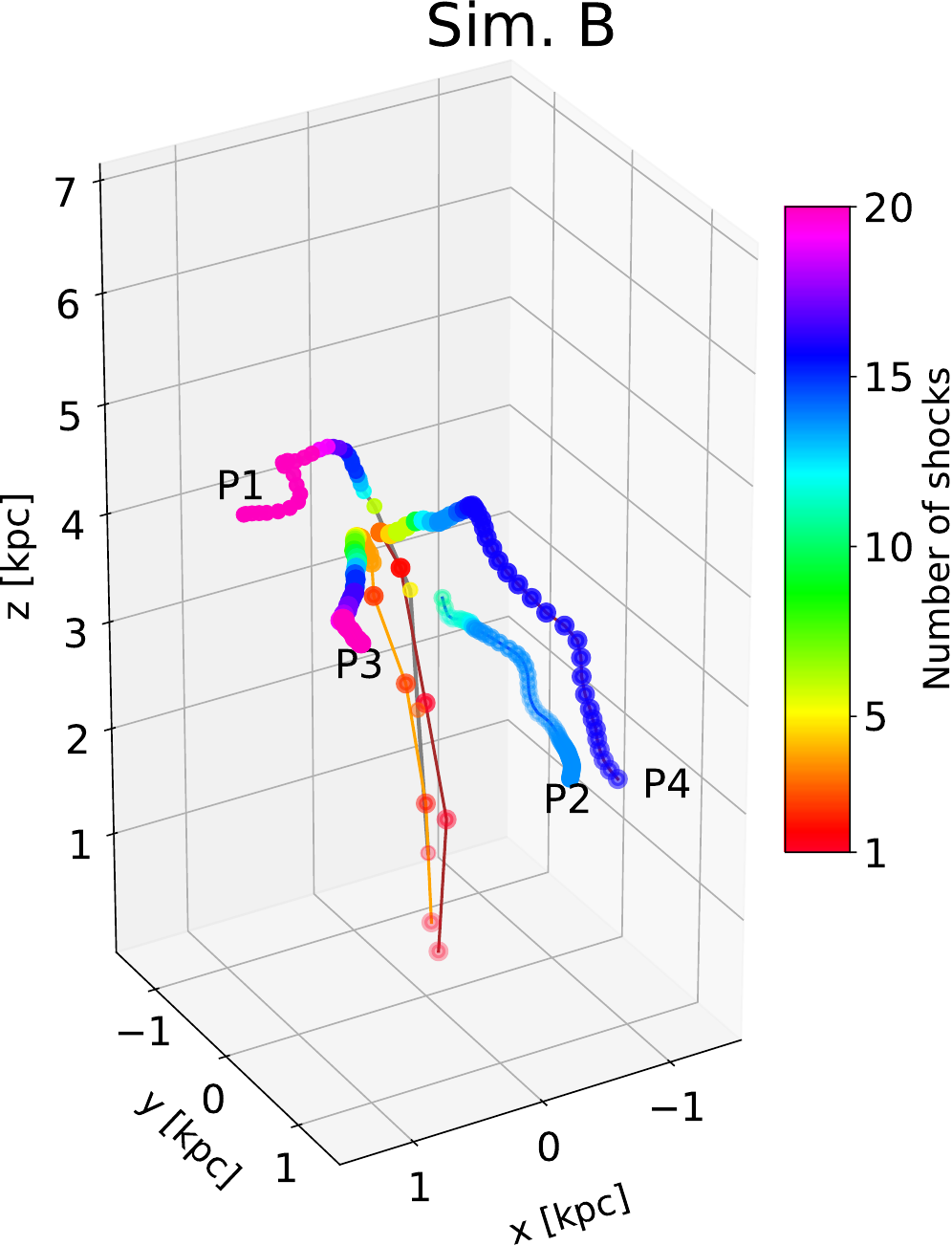}
        \caption{\small \textbf{Left:} Density slice in the $X-Z$ plane for simulation B with $\log(\gamma_{\maxrm})$ of CREs over-plotted, similar to the lower panel of Fig.~\ref{fig.page2D}. \textbf{Right:}  Tracks of 4 representative particles, as also presented in Fig.~\ref{fig.phistSimH}. }
	\label{fig.pgammaSimB}
\end{figure*}
Simulation B with a lower jet power ($P_j\sim 10^{44} \ergs$) shows a uniform distribution of maximum Lorentz factor, as shown in the upper panel of Fig.~\ref{fig.pgammaSimB}. This is in sharp contrast to that of simulation H (Fig.~\ref{fig.page2D}), where the high values of $\gamma_{\maxrm}$ are concentrated at the locations of the strong shocks in the jet axis and the jet-head. The kink instabilities result in the bending of the jet head, ultimately leading to the formation of an extended termination shock spreading horizontally from the main axis of the jet (see top panel in Fig.~\ref{fig.simDshock}). CREs traversing such shock will undergo complex shock crossings. For example, the paths of P1 and P3 are sharply twisted as they encounter the complex shocks at the jet head. They undergo several shock crossings after they exit the jet axis, evidenced from  the rapid change in the color in the lower panel of Fig.~\ref{fig.pgammaSimB}.

The instabilities also strongly decelerate the jet. This results in constant injection of energetic CREs into the slowly inflating cocoon. CREs exiting the jet at different heights, lie close together near the base (e.g. particles P2 and P3). Such CREs are further re-accelerated at other shock surfaces due to internal turbulence and the complex back-flow. For example, particle P3 (right panel of Fig.~\ref{fig.pgammaSimB}) gets further shocked after exiting the jet axis due to internal shocks inside the cocoon.  Thus there is mixing of particles with different shock histories, and the cocoon of a decelerating slowly inflating jet is more evenly distributed with shocked highly energetic particles.

\item{\emph{Simulation D, KH unstable:}}
\begin{figure}
	\centering
	\includegraphics[width = 8. cm, keepaspectratio] 
	{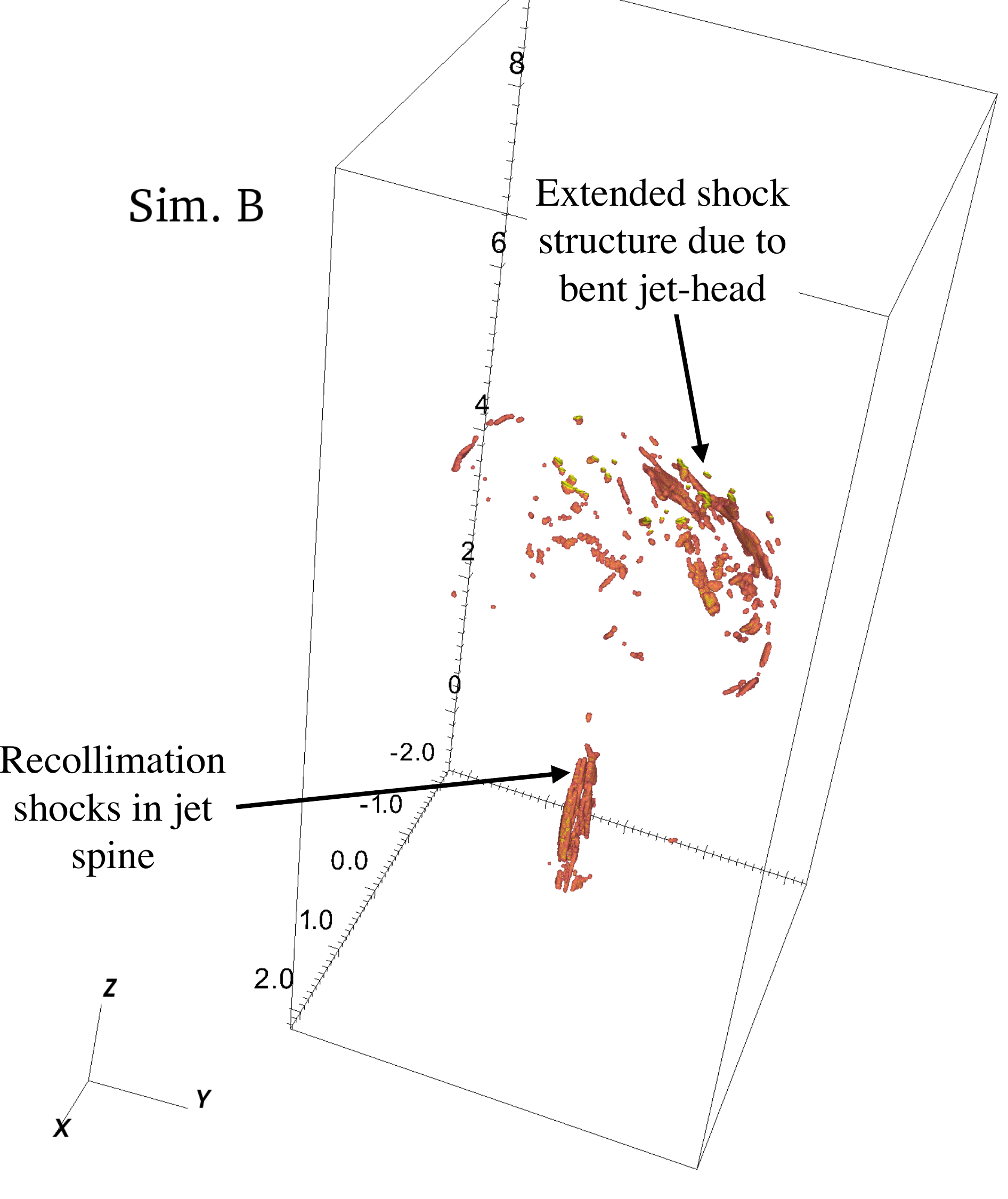}
	\includegraphics[width = 7. cm, keepaspectratio] 
	{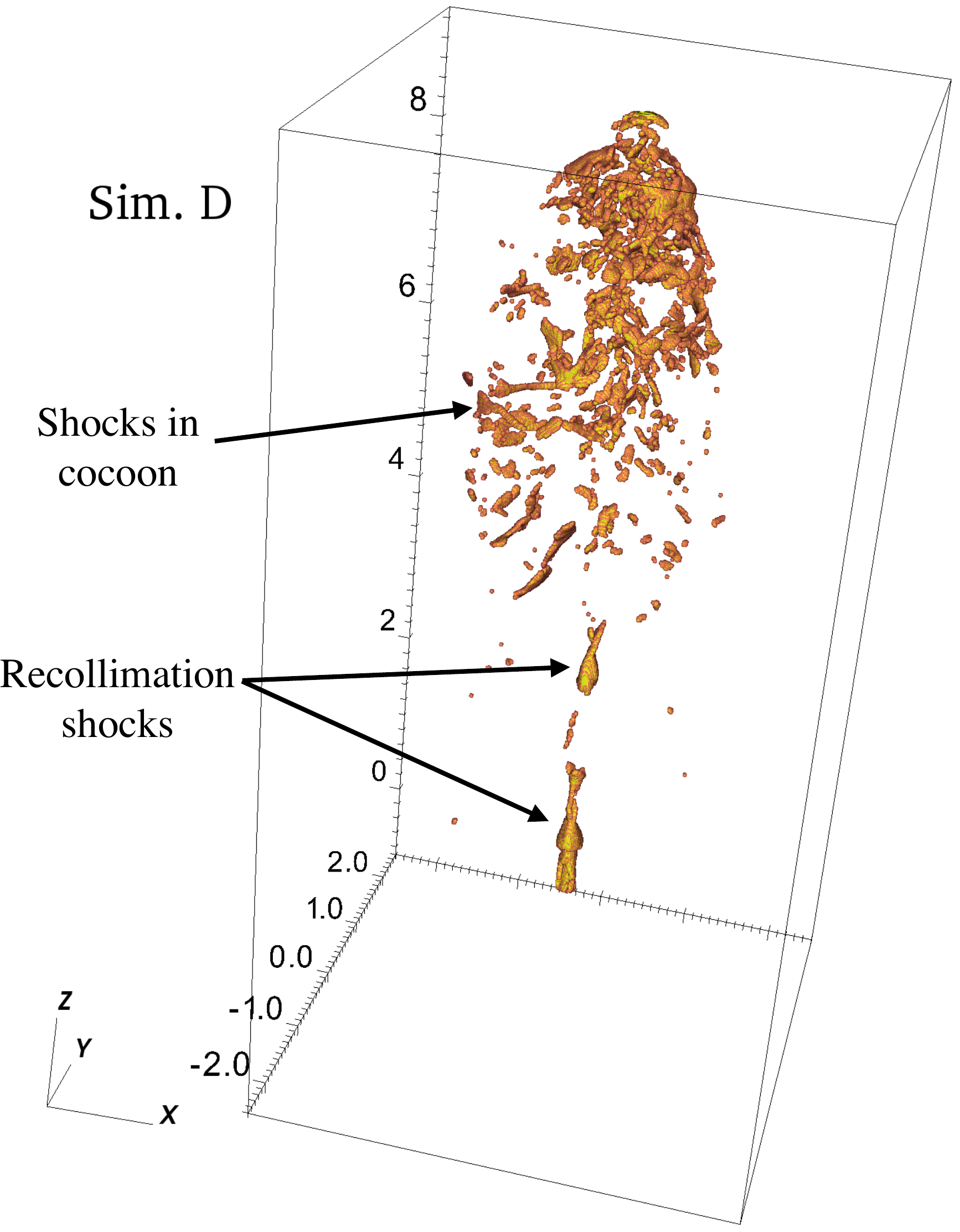}\hspace{-2cm}
        \caption{\small  A 3D volume rendering of shocked surfaces in the cocoon of simulation B (top) and simulation D (bottom). The shocks are identified by the $\Delta p/p > 3$, as described in Sec.~\ref{sec.instablities}. The spatial scale of the axes are in kilo-parsecs. }
	\label{fig.simDshock}
\end{figure}
\begin{figure*}
	\centering
	\includegraphics[width = 4.5 cm, keepaspectratio] 
	{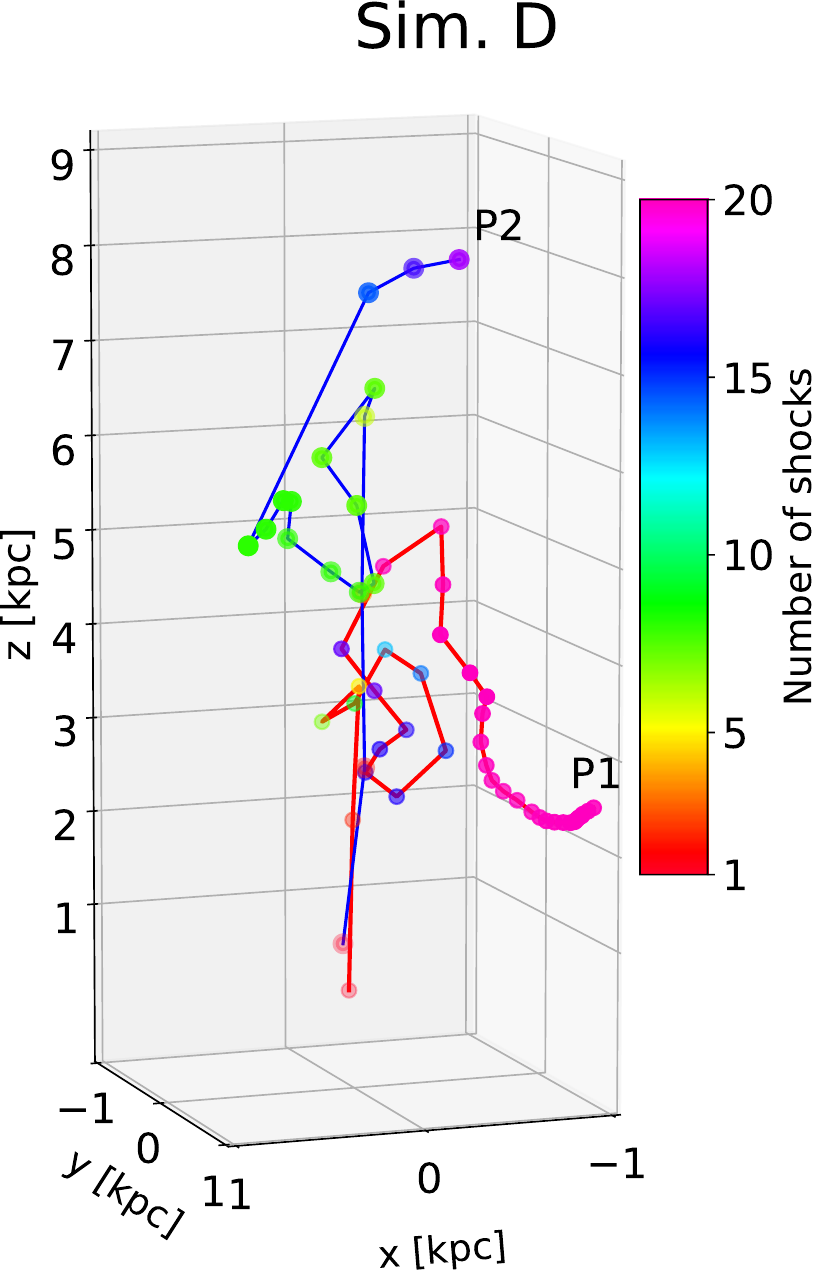}\hspace{-0cm}
	\includegraphics[height = 6.9 cm, keepaspectratio] 
	{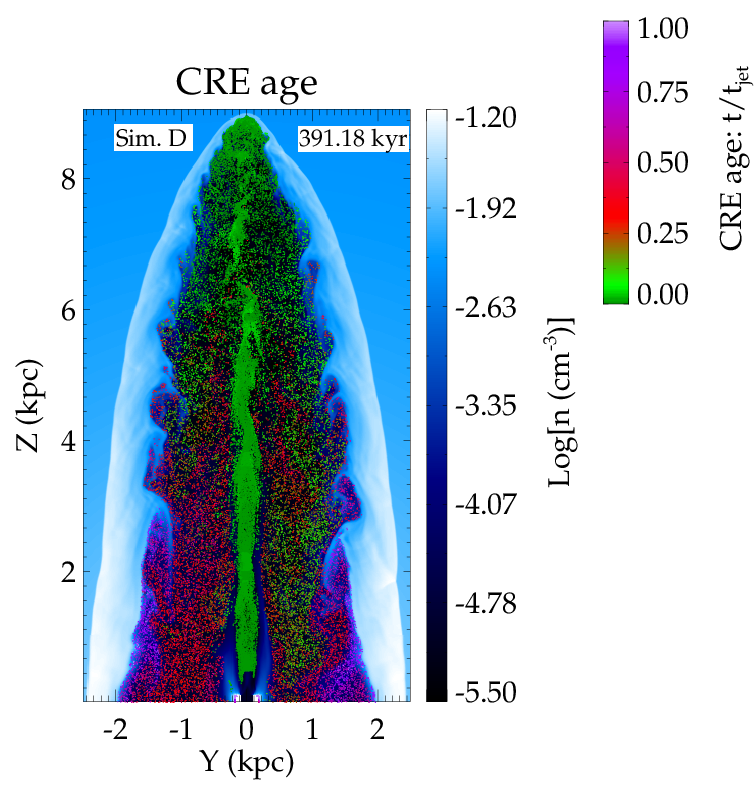}\hspace{0cm}
	\includegraphics[height = 6.9 cm, keepaspectratio] 
	{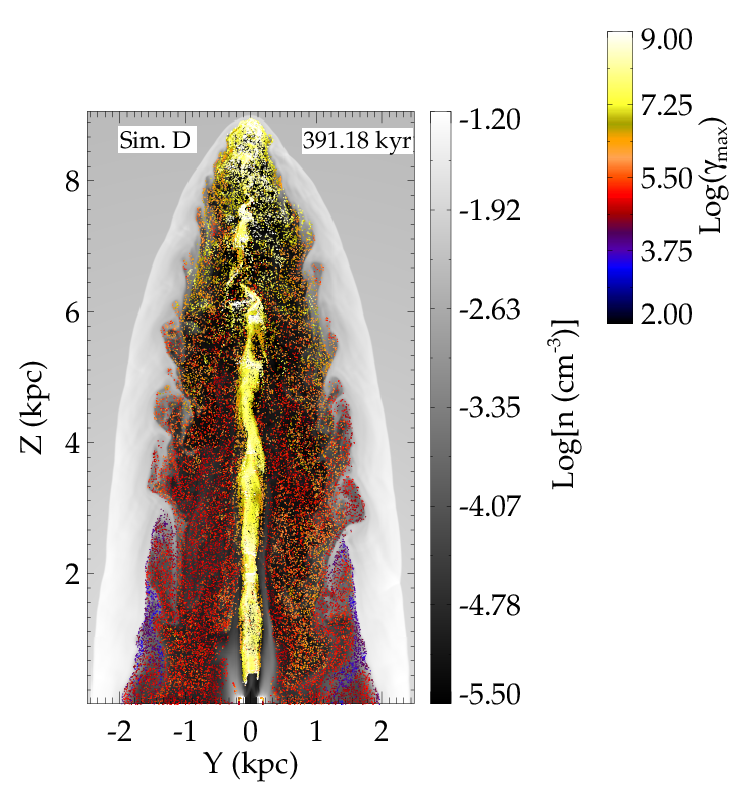}
        \caption{\small \textbf{Left:} Tracks of two representative CREs in simulation D. The track of  P1 is shown by a red-line connecting its locations at different times. Similarly, the track of CRE P2 is represented by the blue line connecting the dots. The color of the dots denotes the number of shock crossings experienced (as also in Fig.~\ref{fig.phistSimH}).  \textbf{Middle:} The density slice for simulation D with particle color showing age since last shock, as in the lower panel of Fig.~\ref{fig.page2D}. \textbf{Right:} Density slice, with particle color showing $\log(\gamma_{\maxrm})$, as in the upper panel of Fig.~\ref{fig.page2D}.  A larger portion of the cocoon has freshly shocked particles due to turbulence generated shocks in the cocoon depicted in the left panel. }
	\label{fig.simD}
\end{figure*}
Simulation D with Kelvin-Helmholtz instabilities has multiple internal shocks inside the cocoon, as shown in the 3D volume rendering of the shocks in Fig.~\ref{fig.simDshock}.  The shock surfaces are identified by computing the maximum pressure difference at each grid point, as defined in \eq{eq.delp}. Similar complex web of shock structures have also been reported in \citet{tregillis01a}. CREs streaming down the backflow will have to cross these shocks and be re-accelerated in the process. CREs shocked multiple times are trapped within this shock layers. 

The above situation is clear by inspecting the complex trajectories of the CREs in the left panel of Fig.~\ref{fig.simD}, where the paths of two representative CREs have been plotted. CRE P1 with the particle trajectory  presented in red lines connecting the locations at different times, is seen to perform a loop at $Z \sim 3$ kpc, before reaching to a maximum height of $Z \sim 5$ kpc and subsequently streaming down. CRE P2 similarly show a complex trajectory (in blue) with a vertical rise up to $Z\sim 6$ kpc, a backward motion thereafter, and again a rise up to $Z\sim 8$ kpc. These vortical motions of the CREs result from the onset of Kelvin-Helmholtz instabilities at the jet-cocoon interface.

What is noteworthy here, is that the cause for the multiple shock crossings in simulation D is fundamentally different than that in simulation B. In simulations B and H, the CREs are shocked multiple times due to the geometry of the shock structure at the jet head, beyond which they stream down along the back flow, with some exceptions. The CREs in simulation D encounter multiple weaker shocks within the body of the cocoon itself as they stream down along the backflow and also at the outer layer of the jet-spine where KH driven vortices arise, leading to turbulence and shocks. 

Thus a more extended region of the cocoon is filled with freshly shocked younger electrons, as shown in middle panel of Fig.~\ref{fig.simD}. Similarly from the right panel of Fig.~\ref{fig.simD} it can be seen that the CREs inside the cocoon are shocked to higher energies, in sharp contrast to that in simulation H (Fig.~\ref{fig.page2D}). This is because the magnetic field in simulation D is almost an order of magnitude lower than that in H, resulting in longer synchrotron cooling time scales (see \eq{eq.gammamax}), allowing efficient acceleration.

\end{itemize}

\subsection{History of the magnetic field, $\gammamax$ and shock compression ratio traced by CREs}\label{sec.gmmhist}
\begin{figure*}
	\centering
	\includegraphics[width = 5.5 cm, keepaspectratio] 
	{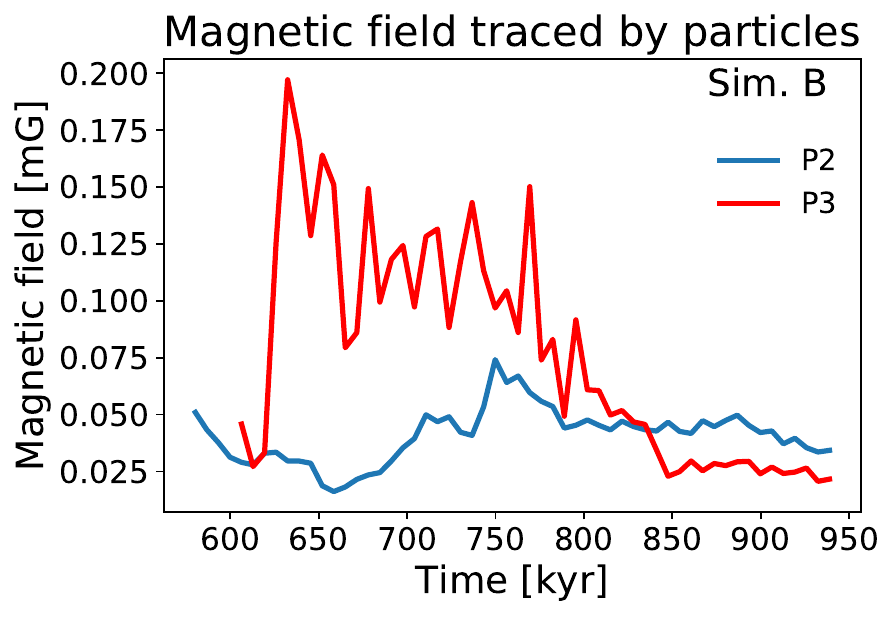}
	\includegraphics[width = 5.5 cm, keepaspectratio] 
	{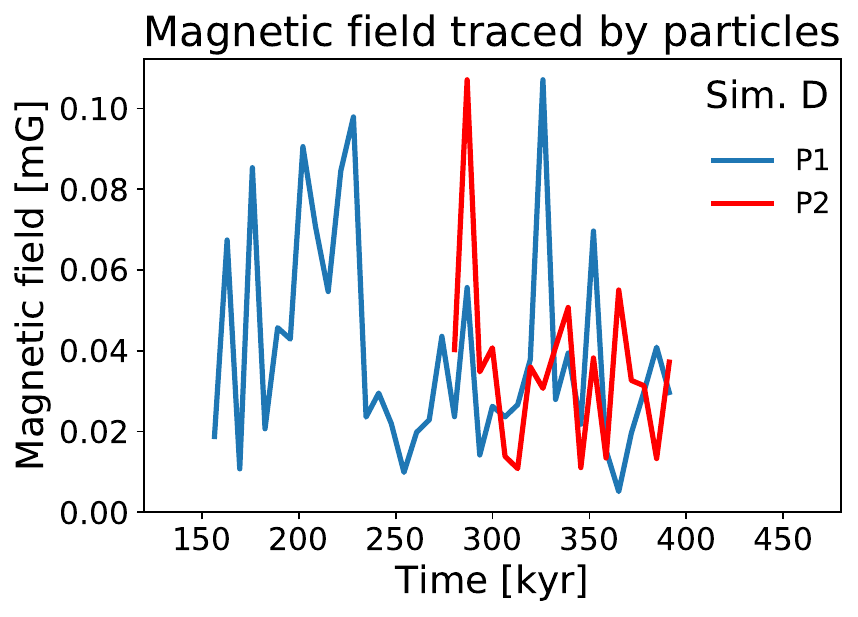}
	\includegraphics[width = 5.5 cm, keepaspectratio] 
	{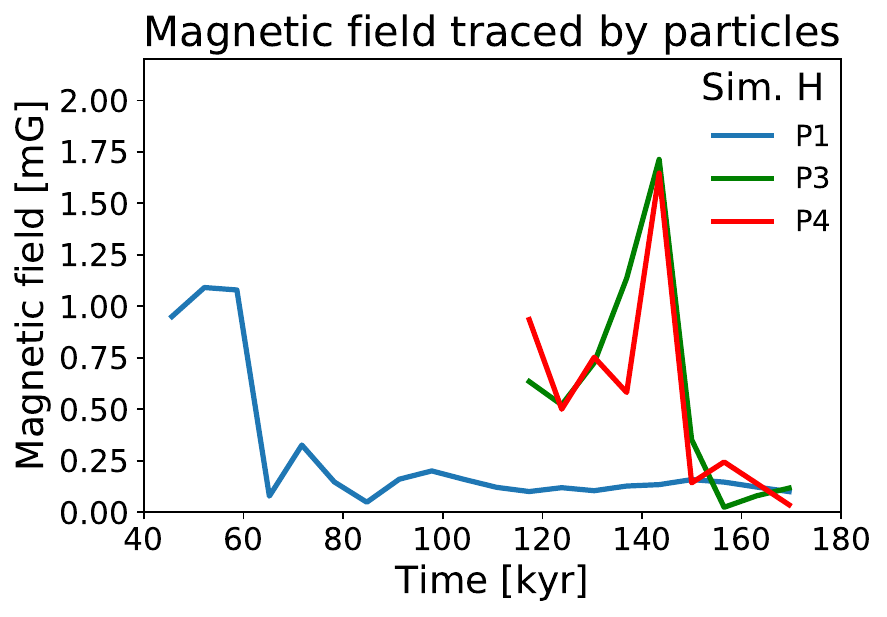}
	\includegraphics[width = 5.5 cm, keepaspectratio] 
	{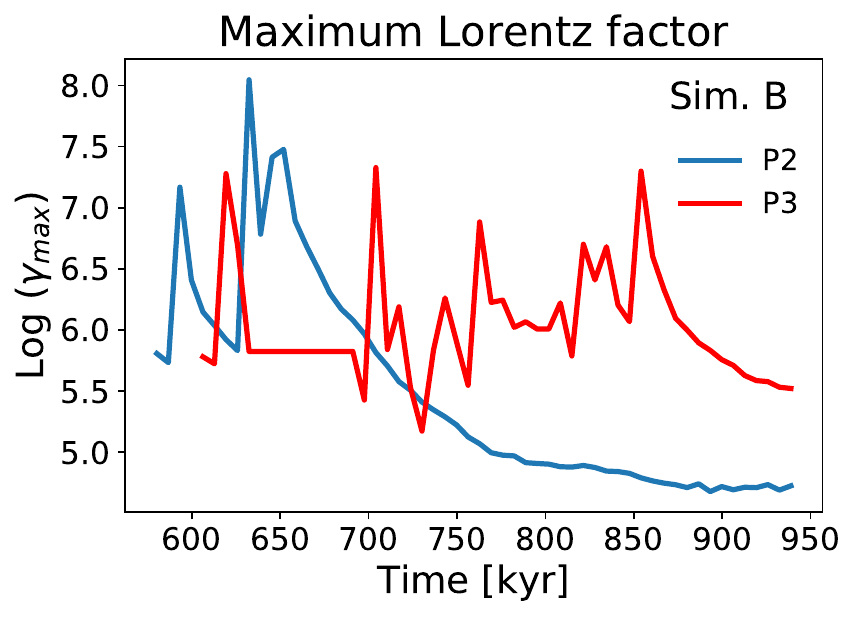}
	\includegraphics[width = 5.5 cm, keepaspectratio] 
	{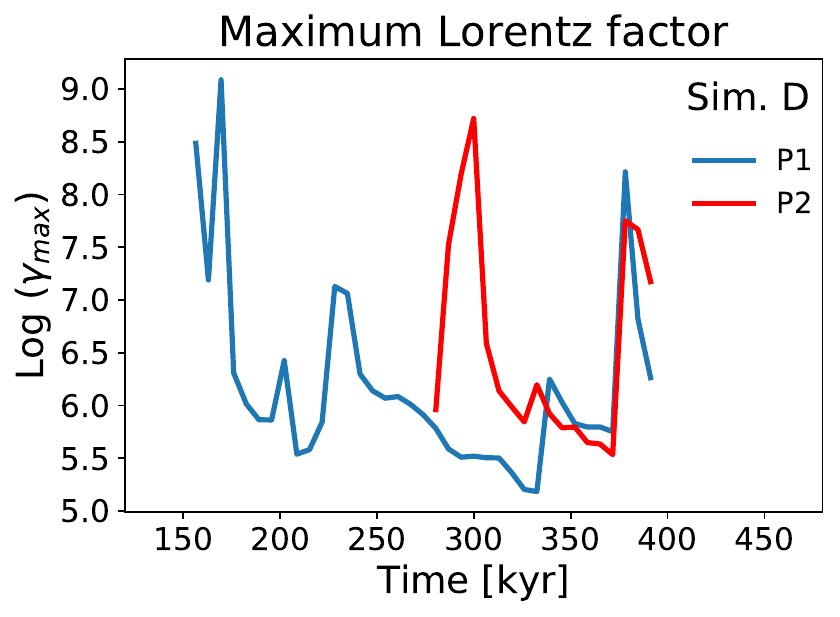}
	\includegraphics[width = 5.5 cm, keepaspectratio] 
	{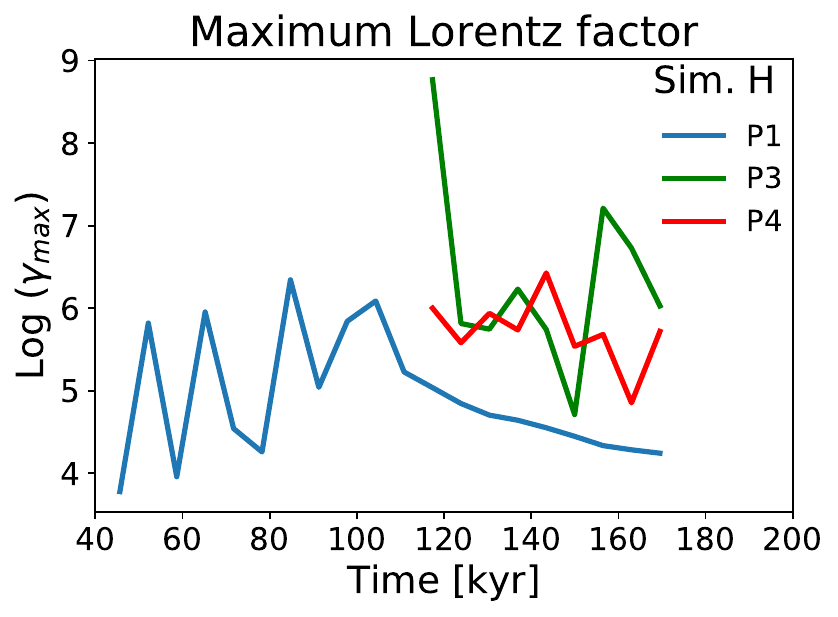}
	\includegraphics[width = 5.5 cm, keepaspectratio] 
	{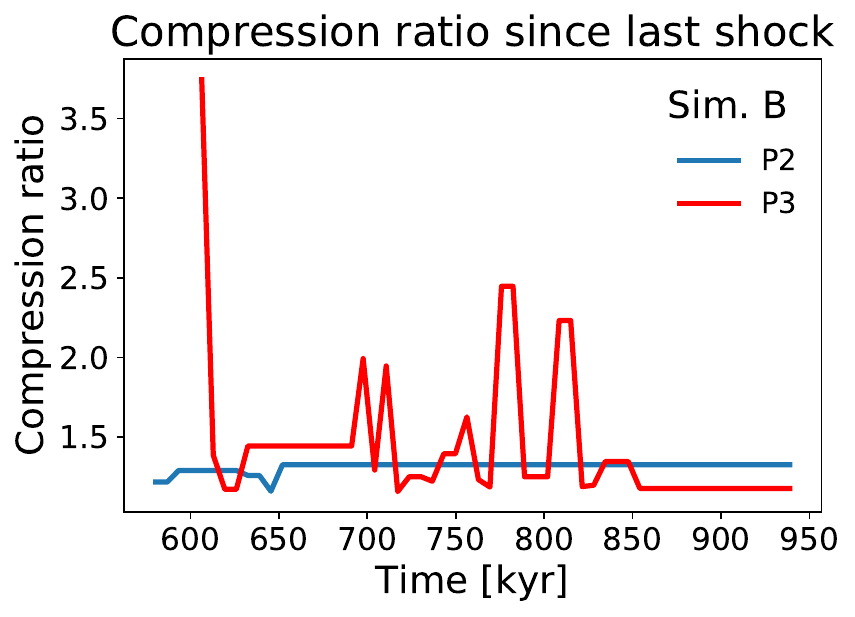}
	\includegraphics[width = 5.5 cm, keepaspectratio] 
	{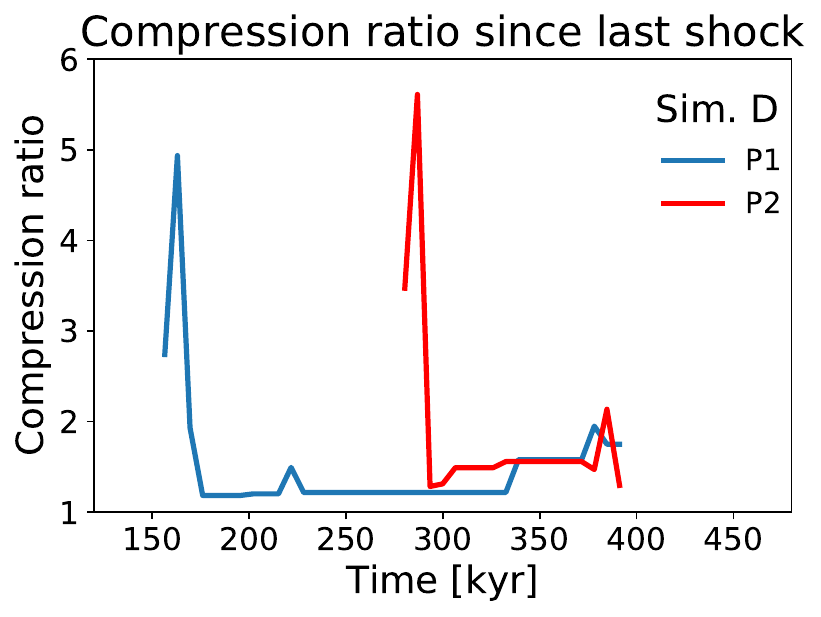}
	\includegraphics[width = 5.5 cm, keepaspectratio] 
	{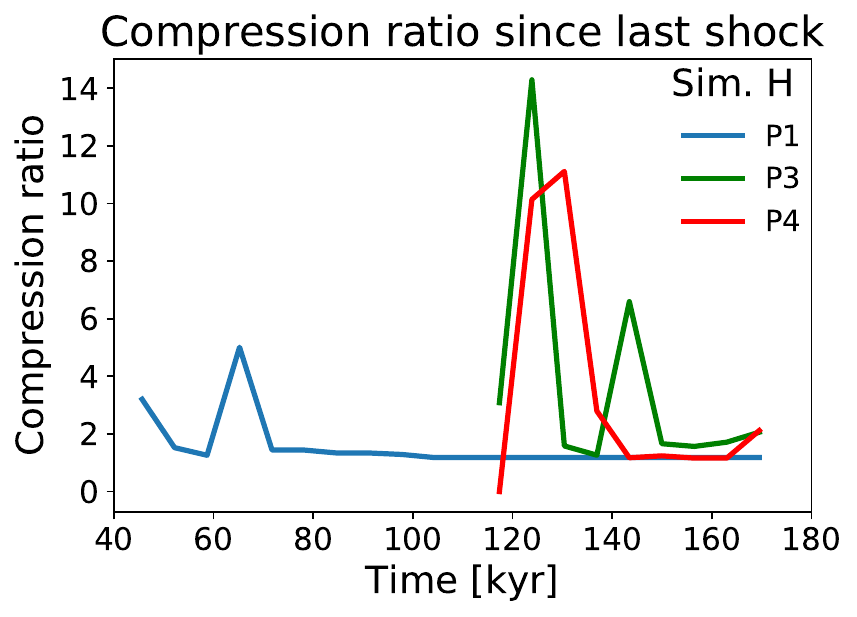}
        \caption{\small The magnetic field (top row), maximum Lorentz factor (middle row) and shock compression ratio since last shock (last panel row) traced by some CREs, whose trajectories are presented in  Fig.~\ref{fig.pgammaSimB}, Fig.~\ref{fig.simD} and Fig.~\ref{fig.phistSimH} for simulations B (left column), D (middle column) and H (right column) respectively.}
	\label{fig.Bhist}
\end{figure*}
MHD instabilities can create turbulence in the jet cocoon, resulting in inhomogeneous distribution of magnetic and velocity fields. Thus, depending on their trajectories in the jet-cocoon, different CRE particles may experience varied  evolutionary history. In Fig.~\ref{fig.Bhist} we present the variation of the magnetic field, the maximum Lorentz factor of the CRE spectrum ($\gamma_{\maxrm}$) and the shock compression ratio of the last shock crossed, for some of the representative particles already shown in Fig.~\ref{fig.phistSimH}, Fig.~\ref{fig.pgammaSimB} and Fig.~\ref{fig.simD}. The variation of these quantities along the CRE trajectory will leave their imprint on the CRE energy spectrum. 
\begin{itemize}
\item \emph{Simulation B:}
The left column of Fig.~\ref{fig.Bhist} shows the evolution of physical properties for particles P2 and P3, whose trajectories have been presented in Fig.~\ref{fig.pgammaSimB}. We firstly notice that although the two particles have similar trajectories, they encounter very different magnetic fields. Particle P3 shows a sharp increase to about $\sim 0.2$ mG, followed by a decline, and similar subsequent spikes, although of lower magnitudes. The rapid increase in the magnetic fields correlate well with changes in the shock-compression ratio, indicating that such sudden increase results from the CRE traversing different shocks during its trajectory. The $\gamma_{\maxrm}$ also shows similar correlated increase to high values, followed by an exponential decline after the last shock. It is interesting to note that since its injection after $\sim 650$ kyr of the simulation run time, CRE P3 experiences multiple shocks for $\sim 200$ kyr, before entering a steady quiescent backflow without shocks. This is due to the complex shock structure created by the kink unstable bending jet-head.

CRE P2 however, experiences a very different evolution, with a steady magnetic field fluctuating about a mean of $\sim 0.04$ mG. The $\gamma_{\maxrm}$ shows some initial increase as the CRE travels through the jet-axis and reaches the complex shock at the jet-head, as also corroborated from the initial changes in compression ratio. However, post jet exit, the CRE streams through the backflow with an exponential decay of $\gamma_{\maxrm}$. The evolution of P2 is thus along the lines of the standard expectation of CRE evolution proposed in traditional analytical models \citep{jaffe73a,turner15a}, as also depicted in Fig.~\ref{fig.jetCartoon}. However, CRE P3 with multiple shock crossings spread over a long duration, is a result of the local inhomogeneities due to non-linear MHD. 

Another point to note is that, the high $\gamma_{\maxrm}$ of CRE P3 coincide with high values of the compression ratio ($r > 1.5$). This is expected, as strong shocks of higher compression ratio ($r$), will give rise to higher values of $\gamma_{\maxrm}$ for the same strength of magnetic field, as can be seen in \eq{eq.gammamax} and \eq{eq.acc}. However, for CRE P2, the initial rise in $\gamma_{\maxrm}$ corresponds to crossings of weak shock ($r \lesssim 1.3$), and also lower magnetic field strengths than P3 by nearly an order of magnitude. As can again been seen from \eq{eq.gammamax}, lower values of magnetic field strength at shocks will also result in an increase in $\gamma_{\maxrm}$ due to lower synchrotron cooling times. Thus these two CREs demonstrate well how CREs can experience different physical conditions during their trajectory, and how they differently affect their spectra.

\item \emph{Simulation D:}
For simulation D (middle column of Fig.~\ref{fig.Bhist}), we present the results for CREs P1 and P2 labelled in Fig.~\ref{fig.simD}, which have very different spatial tracks, as discussed earlier in Sec.~\ref{sec.instablities}. The turbulent magnetic field fluctuating on smaller length scales than other simulations results in the CREs experiencing a magnetic field fluctuating about a mean value ($\sim 0.04$ mG) as well. The $\gammamax$ however shows sharp increase followed by phases of decline. The increase in $\gammamax$ correlates with an increase in the compression ratio, indicating that the particles have traversed through strong shocks with compression ratios $r \gtrsim 2-4$. The multiple peaks in the $\gammamax$ at different times result from re-acceleration at internal shocks during their motion within the cocoon. The low mean injected magnetic field in this simulation, also contribute to the high values of $\gammamax$.

\item \emph{Simulation H:}
For simulation H, we present the results for 3 CREs. CRE P1 exits the jet at $Z\sim 3$ kpc and proceeds laterally onwards into the cocoon. CREs P3 and P4 injected at later times exit at a higher height ($Z\sim 9$ kpc). As CRE P1 proceeds through the jet axis, it experiences an initial rise of magnetic field along its trajectory, followed by a lower value of $\sim 0.1$ mG. CREs P3 and P4 having been injected at similar times, follow similar trajectories and show a similar nature of the time evolution of the magnetic field and shock compression ratio. The initial $\gamma_{\maxrm}$ of P3 is higher as it likely passes through a stronger recollimation shock after injection, than P4. 

Although the CREs experience  stronger shocks than simulation D,  the $\gamma_{\maxrm}$ is not much higher than that in simulation D. The stronger magnetic field in simulation H, causes faster radiative losses of the CREs reducing the efficiency of shock acceleration. 
\end{itemize}

\subsection{Multiple internal shocks}\label{sec.multishock}
\begin{figure}
	\centering
	\includegraphics[width = 8 cm, keepaspectratio] 
	{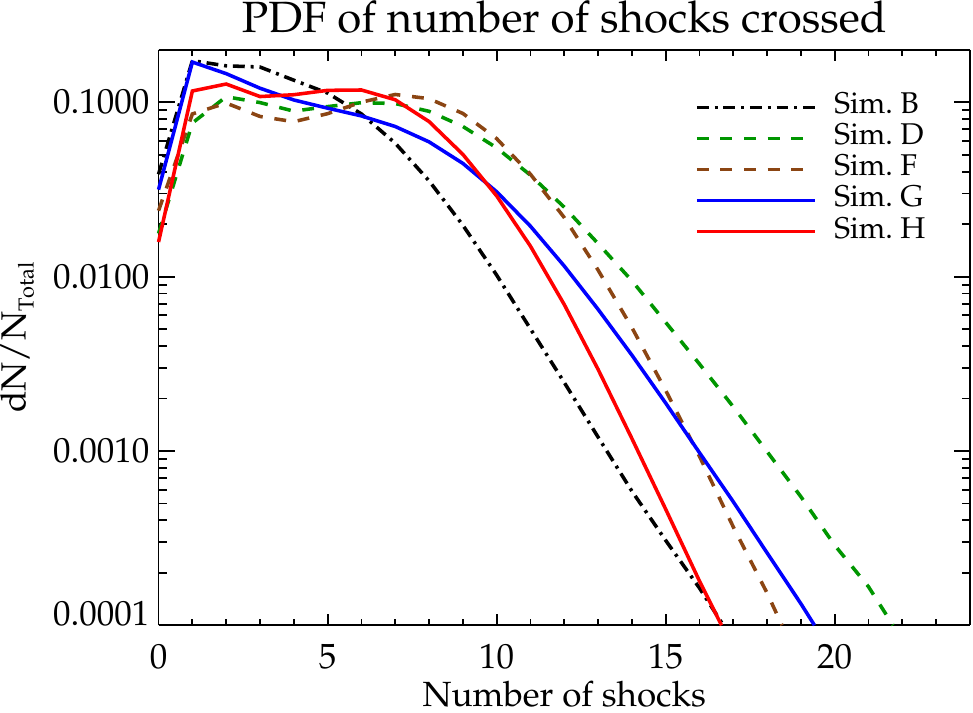}
        \caption{\small The distribution of number of shocks traversed by the particles during their lifetime for simulations B, D, F, G and H. Stable jets in F and H have steep tail due to lower shock crossings. Simulation D with Kelvin-Helmholtz modes has an extended tail indicating many crossings.}
	\label{fig.nshockpdf}
\end{figure}
As stated earlier in previous sections, MHD instabilities create complex shock structures inside the jet cocoon where CREs can be re-accelerated. Fig.~\ref{fig.nshockpdf} shows the probability distribution function (hereafter PDF)  of the number of shocks encountered by CREs in different simulations. Simulations with similar power are plotted with same linestyles although different colours viz. simulations D (green) and F (brown) with $P_{\rm jet} \sim 10^{45}\ergs$ in dashed, and simulations G (blue) and H (red) with $P_{\rm jet} \sim 10^{46}\ergs$ in solid. Kelvin-Helmholtz instabilities in simulation D result in higher number of shock crossing with the PDF showing a very extended tail than that in others. Similarly, simulation G which has more internal structures and shocks due to higher internal sound speed has a slightly extended PDF than the stable jet in simulation H. Simulations F and H show a slight hint of bi-modality. This results from joint contributions of the CREs in the jet spine which are shocked only at a few recollimation shocks and the CREs in the cocoon which have been shocked many times at the jet head or the internal shocks in the cocoon.

\subsection{Maximum Lorentz factor of CREs}
\subsubsection{Distribution of $\gamma_{\maxrm}$ at different heights}\label{sec.gammamax}
\begin{figure*}
	\centering
	\includegraphics[width = 15 cm, keepaspectratio] 
	{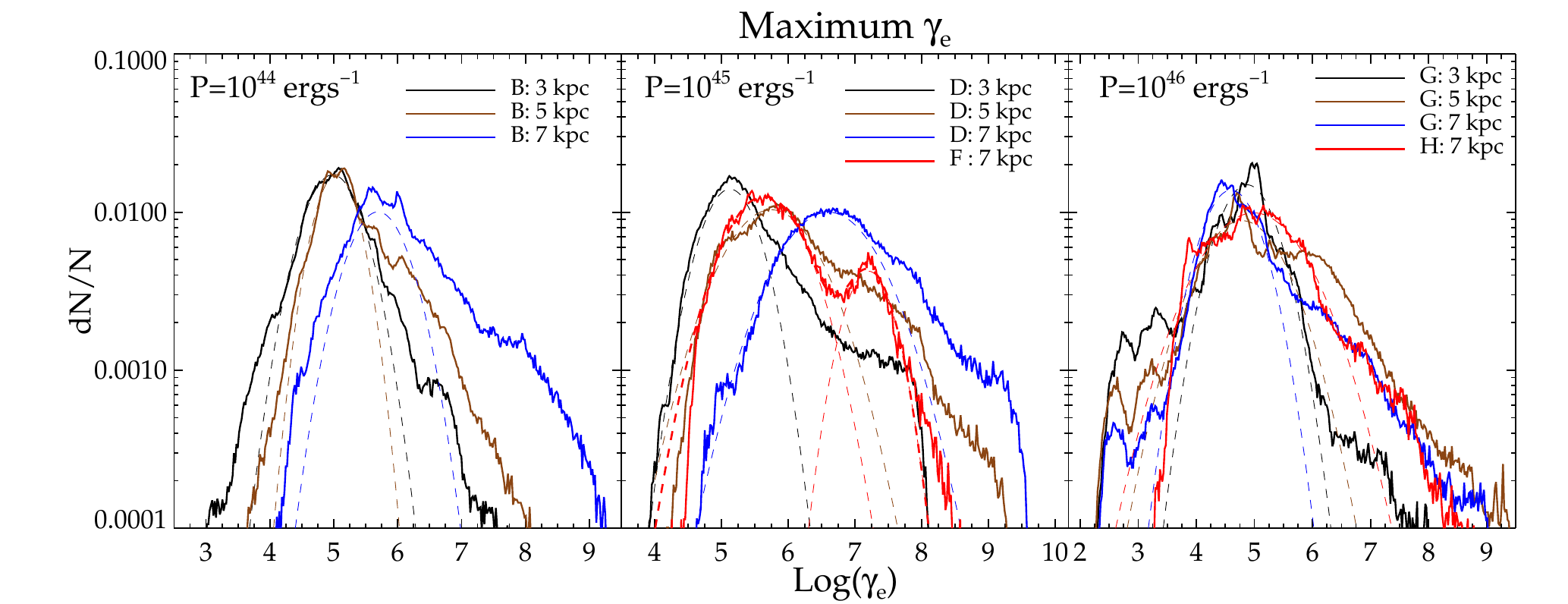} 
        \caption{\small The  distribution of $\gamma$ for different simulations, at three different heights, following the convention of Fig.~\ref{fig.nshockpdf}. The three panels correspond to simulations of in 3 ranges of power. The PDFs in general show a peak (which is well described by a log-normal distribution shown in dashed lines), followed by an extended tail to high $\gamma$.  }
	\label{fig.emaxpdf}
\end{figure*}
In this section we discuss the effect of multiple shock encounters on the maximum Lorentz factor of the CREs ($\gamma_{\maxrm}$), which is defined by \eq{eq.gammamax}.  Fig.~\ref{fig.emaxpdf} shows the PDF of the $\gamma_{\maxrm}$ at three different heights, corresponding to the jet-head ($\sim 7$ kpc, in blue), the middle zone ($\sim 5$ kpc, in brown) and the base of the cocoon ($\sim 3$ kpc, in black), for different simulations. The PDFs at different heights are instructive to understand the evolution of the CRE spectra, as they are shocked at the jet-head and later again while crossing weaker shocks in the back-flow.

Firstly, it is evident that the PDFs have a more extended distribution ($\gamma_{\maxrm} \sim 10^9$) closer to the jet head ($\sim 7$ kpc), where they encounter the strong terminal shock. The PDFs in the mid-planes for simulation B have a lower value of maximum Lorentz factor ($\gamma_{\rm max} \lesssim 10^8$) as they consist of particles that have cooled due to radiative losses. However, in simulations D and G,  $\gamma_{\rm max}$ in the middle zone extends up to $\gamma_{\rm max} \sim 10^9$ as  particles get re-accelerated by internal shocks inside the cocoon. 

The PDFs have a general nature of a peak at $\gamma_{\rm max} \sim 10^5$, followed by an extended tail. The peak and the lower end of the PDF is often well described by a log-normal distribution, as shown by the dotted lines in Fig.~\ref{fig.emaxpdf}. The peak of the lognormal shows a decreasing trend with distance from the jet-head, indicating radiative losses. The tail of the PDF extending beyond the log-normal to high Lorentz factors ($\gamma_{\rm max} \sim 10^8 - 10^9$), comprises of freshly shocked electrons. Such particles thus form a different population of highly energetic, freshly shocked electrons. The PDF at $\sim 7$ kpc for simulation F (middle panel, in red) indeed shows a bi-modal distribution with the higher peak also described by a log-normal like distribution.

\subsubsection{Multiple shocked components}\label{sec.gammamax}
\begin{figure*}
	\centering
	\includegraphics[height = 3.3 cm, keepaspectratio] 
	{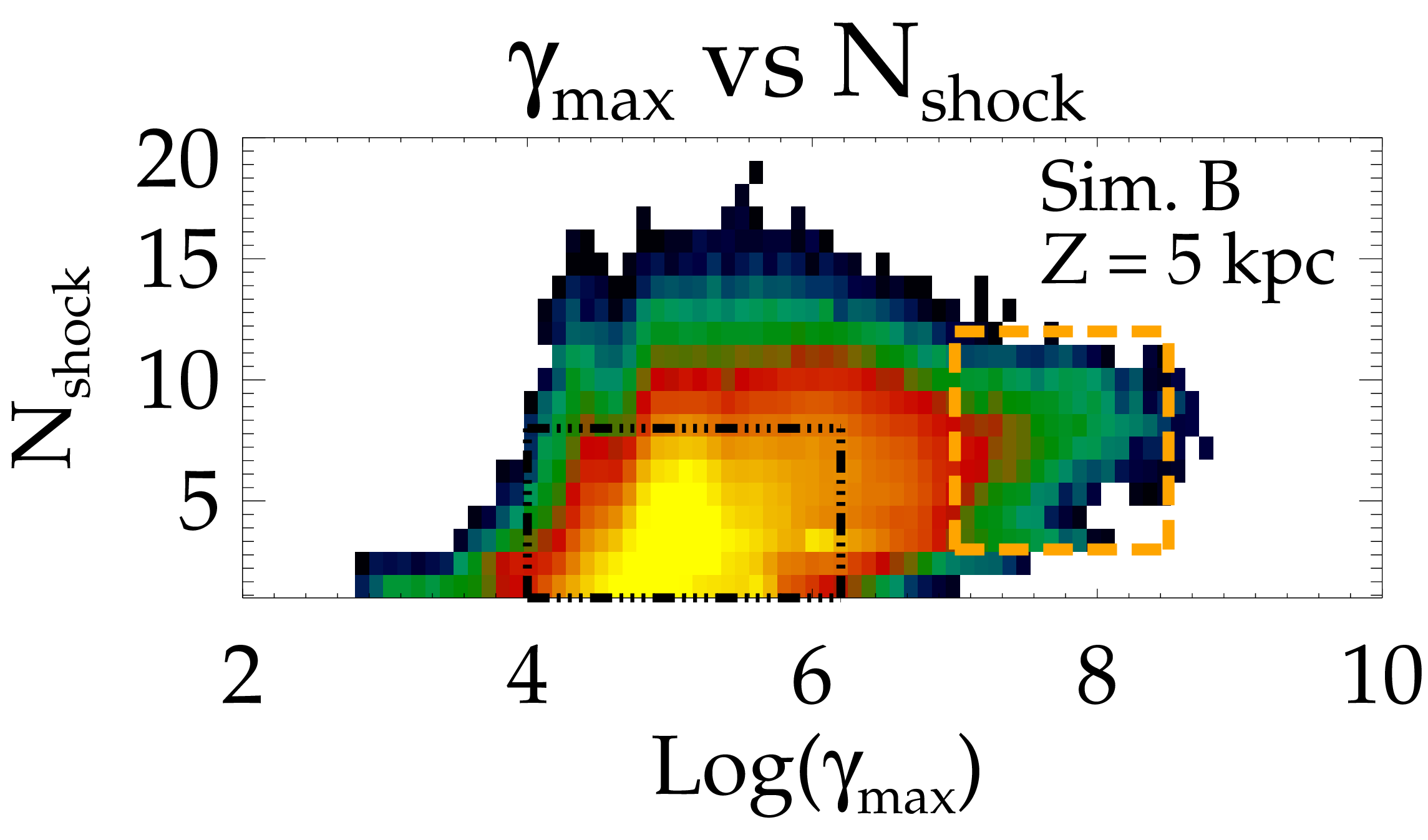}
	\includegraphics[height = 3.3 cm, keepaspectratio] 
	{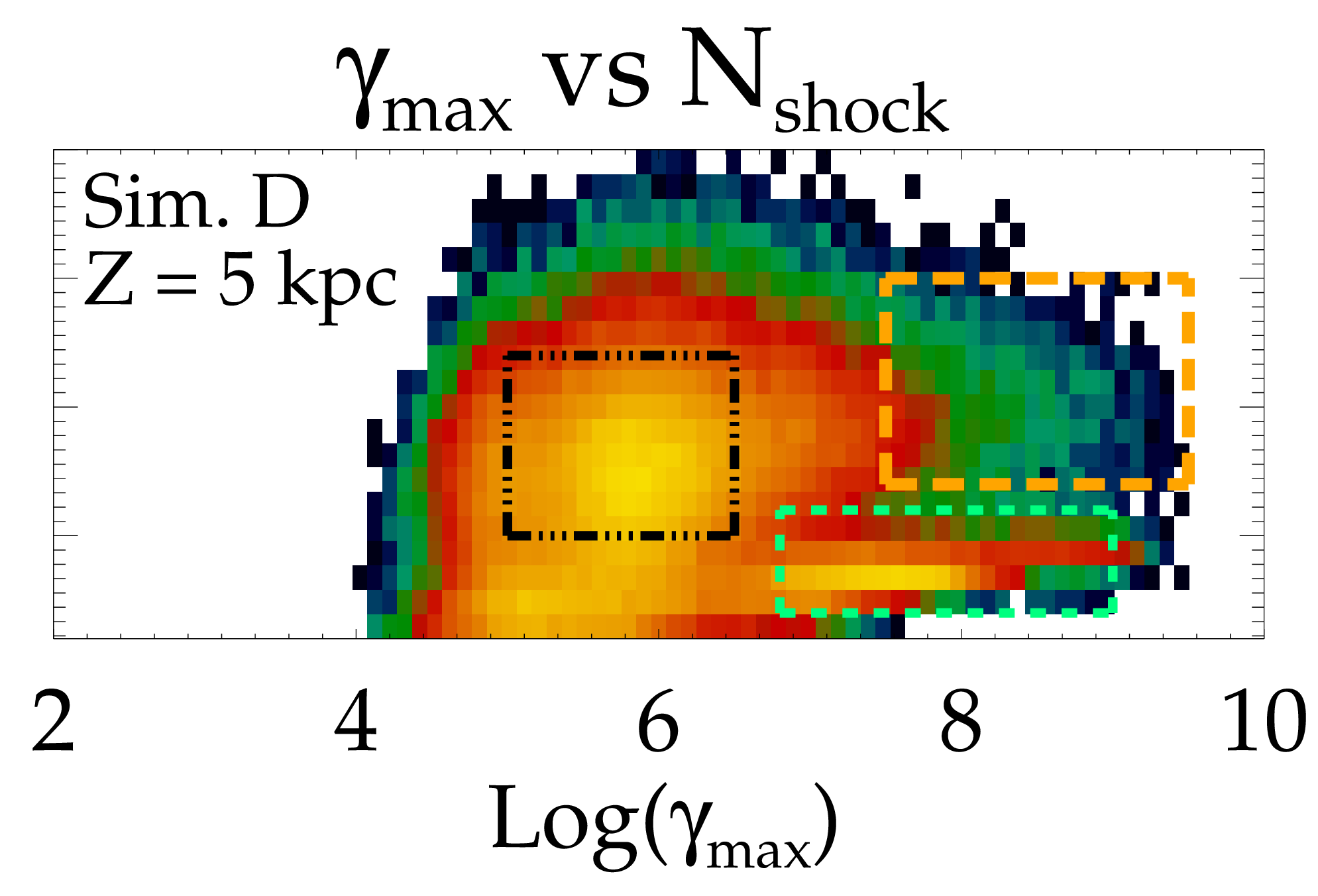}
	\includegraphics[height = 3.3 cm, keepaspectratio] 
	{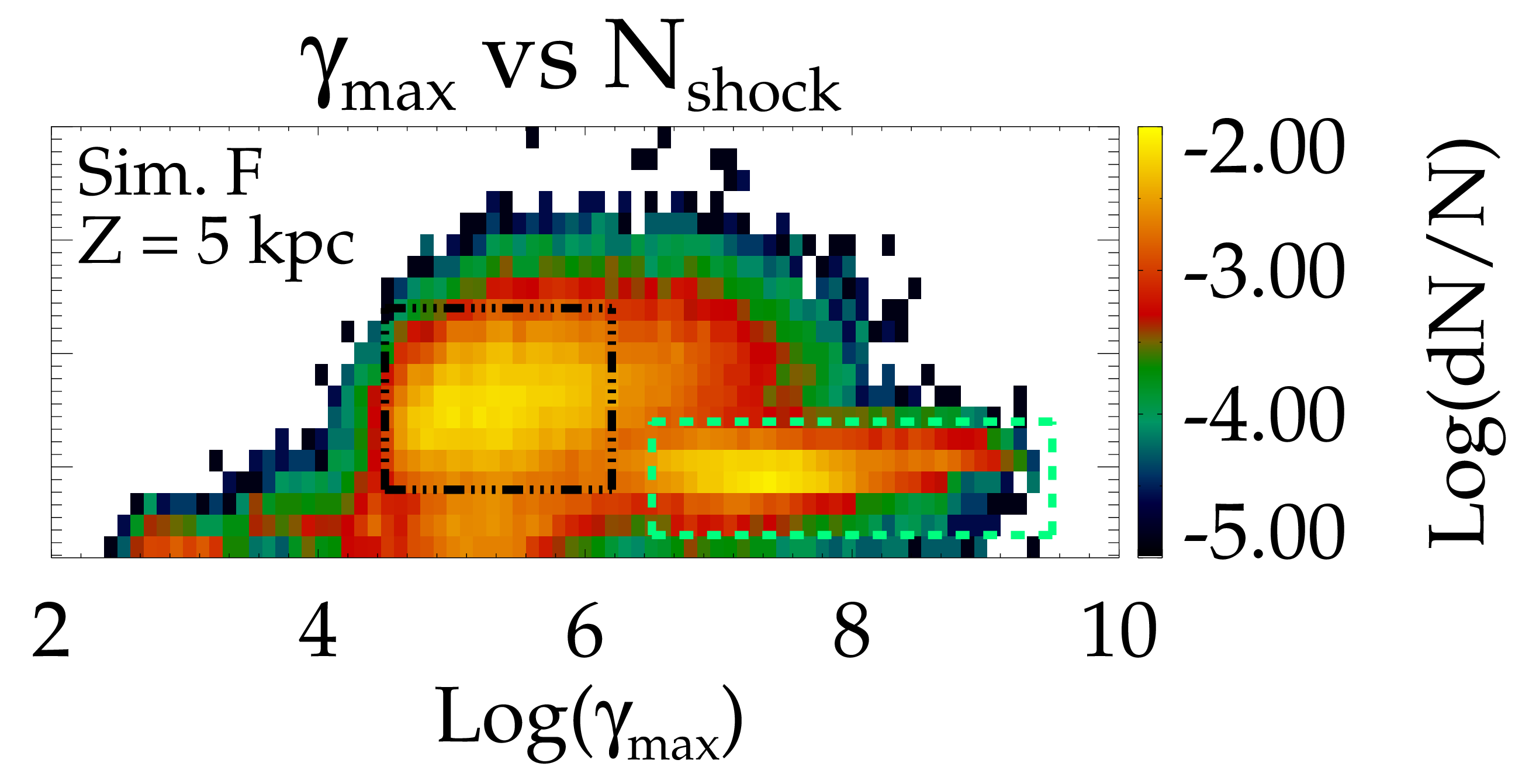} 
        \caption{\small 2D PDF of $\gamma_{\maxrm}$ vs number of shocks crossed ($\mbox{N}_{\rm shock}$) by a CRE macro-particle at a height $z = 5$ kpc at the end of the simulation. Three different CRE populations have been highlighted in coloured boxes. 1) \textit{Pop. I} in the black box:  Older CREs with decayed spectra. 2) \textit{Pop. II} in green box:  Freshly shocked CREs with high energies that lie in the jet spine. 3) \textit{Pop. III} in orange box: re-accelerated CREs that have been shocked many times. }
	\label{fig.2Dpdf}
\end{figure*}
The onset of MHD instabilities creates different population of CREs with different shock histories and energetics. This can be seen from the 2D PDFs of the number of shock crossings of the CREs and the $\gamma_{\maxrm}$ in  Fig.~\ref{fig.2Dpdf}. The PDFs have been made at a height of $z = 5\pm0.2$ kpc, which is approximately half the jet-height at the end of the simulation. The half height was selected so that CREs have enough time to settle into the backflow inside the cocoon, after exiting jet.

We can see that in general three zones can be identified, for three different populations of CREs:
\begin{itemize}
\item \emph{Pop. I:} They are represented approximately by the blackbox in dash-dotted contours, over-plotted on the PDF maps in Fig.~\ref{fig.2Dpdf}. Such CREs have experienced 5-10 shock crossing (lower for simulation B) and have $\gamma_{\maxrm} \sim 10^4 - 10^6$, peaking around $\gamma_{\maxrm} \sim 10^5$. This is typical for CREs whose higher energy part of the spectra is decaying due to radiative losses. The mean spread of $\gammamax$ of these CREs corresponds well with the peak of the log-normal distribution at $\sim 5$ kpc in Fig.~\ref{fig.emaxpdf} discusses earlier. Thus these CREs represent the bulk of the non-thermal electrons in the cocoon whose spectrum follow the standard evolutionary scenario of being accelerated at the jet-head and followed by radiative cooling, as shown in the cartoon in Fig.~\ref{fig.jetCartoon}.
\item \emph{Pop. II:} There is a second population of CREs, especially for simulation D and F, with high energies ($\gamma_{\maxrm} \sim 10^7-10^9$), but less number of shock crossings ($\mbox{N}_{\rm shock} \sim 2-5$). They are highlighted by a green box with dashed contours in the middle and right panels of Fig.~\ref{fig.2Dpdf}. These are CREs in the jet spine, which have been energised by recollimation shocks, and hence the high $\gamma_{\maxrm}$. Since, there are only a few of recollimation shocks in the jet, they have undergone only a handful of shock-crossings.  
\item \emph{Pop. III:} A third population of CREs can be identified in simulations B and D, which have high $\gamma_{\maxrm} \sim 10^7 - 10^9$ and multiple shock crossings ($N_{\rm shock} > 5$), denoted by the orange box in the left and middle figures of the top panel in Fig.~\ref{fig.2Dpdf}. These CRE have crossed multiple shocks either in the complex structure at the jet-head (as in simulation B) or internal shocks inside the cocoon (e.g. simulation D). This is a signature of jets with active MHD instabilities, and is distinctly absent in simulation F, which is stable to both kink and KH modes, as also reported recently in \citet{borse21a}. In simulation F, after crossing the jet-head, the CREs cool down in the backflow, without being re-accelerated, unlike the CREs in \textit{Pop. III} category.
\end{itemize}

\subsection{Distribution of CRE ages at a given height}\label{sec.tage}
\begin{figure*}
	\centering
	\includegraphics[height = 6.1 cm, keepaspectratio]{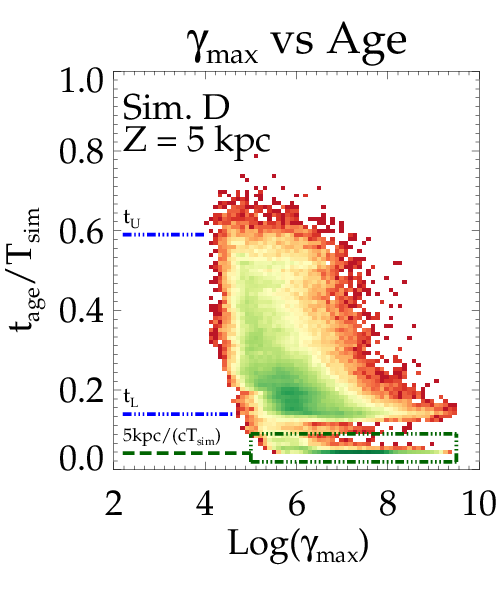}
	\includegraphics[height = 6.1 cm, keepaspectratio]{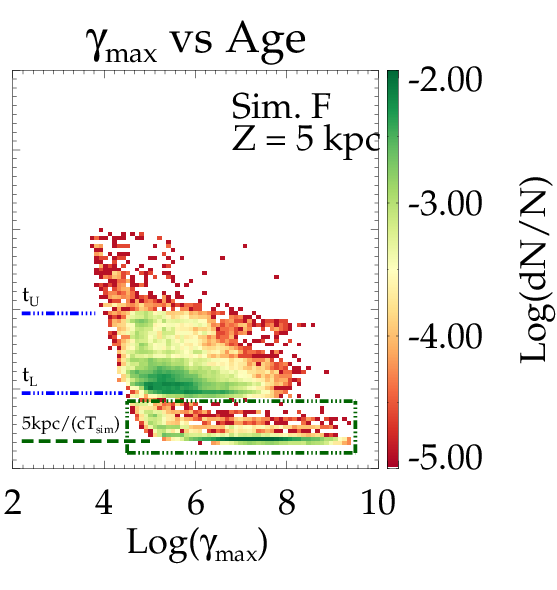}
	\includegraphics[height = 4.5 cm, keepaspectratio]{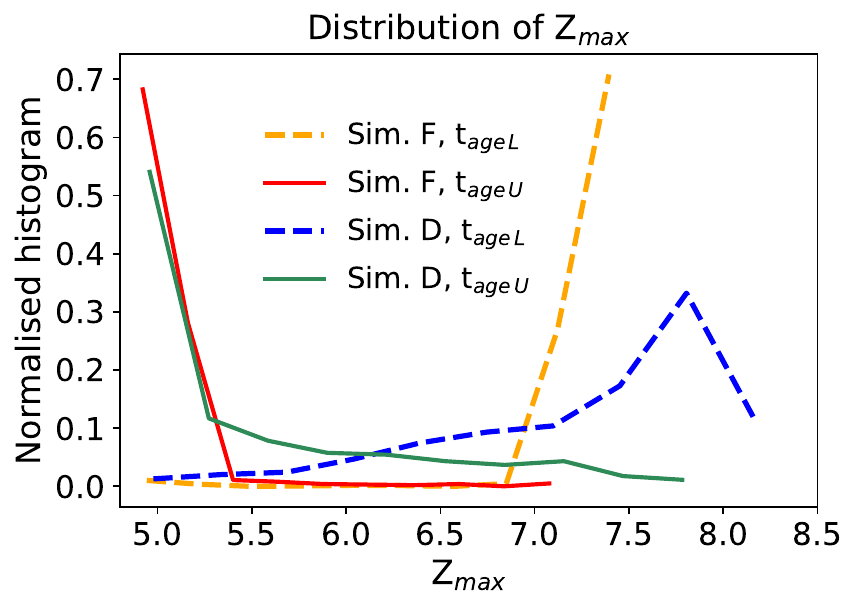}
        \caption{\small \textbf{Left \& Middle:} 2D PDF of $\gamma_{\maxrm}$ vs time elapsed since a CRE is injected ($t_{\rm age}$) representing the age of a CRE in the simulation, normalised to the total simulation time ($T_{\rm sim}$). The \textit{green box} denotes CREs inside the jet.  The green dashed line denotes the time to reach $z = 5$ kpc with a speed $c$, normalised to the simulation time.  The blue lines, marked by $t_L$ and $t_U$ denote the extent of the distribution of ages within the cocoon that have exited the jet. See text in Sec.~\ref{sec.tage} for further details. \textbf{Right:} The distribution of maximum height ($Z_{\rm max}$) reached by particles with the ages $t_{L}$ and $t_{U}$ in the previous panels.
 }
	\label{fig.2DpdfTage}
\end{figure*}
\begin{figure}
	\centering
	\includegraphics[height = 7.5 cm, keepaspectratio]{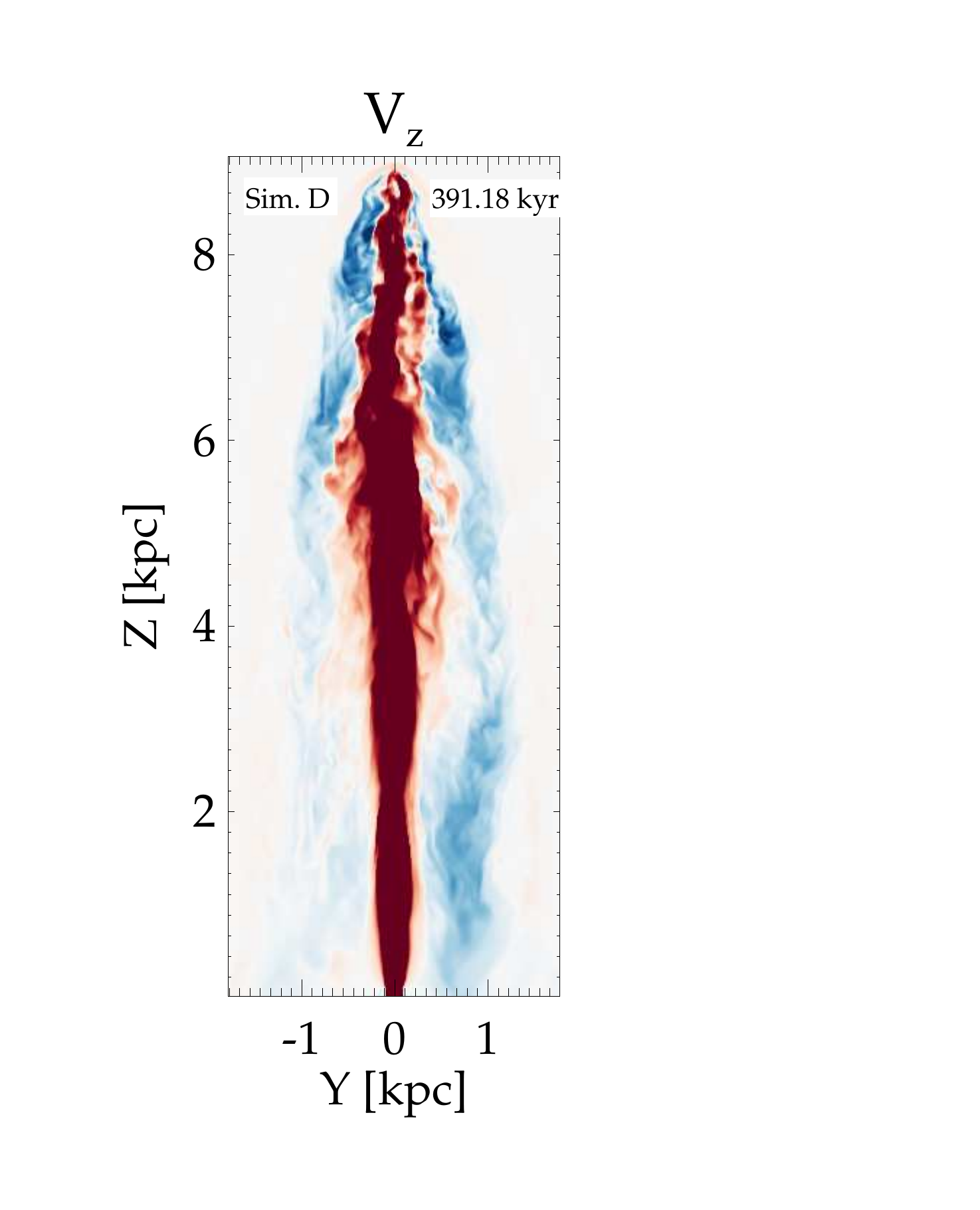}\hspace{-4cm}
	\includegraphics[height = 7.5 cm, keepaspectratio]{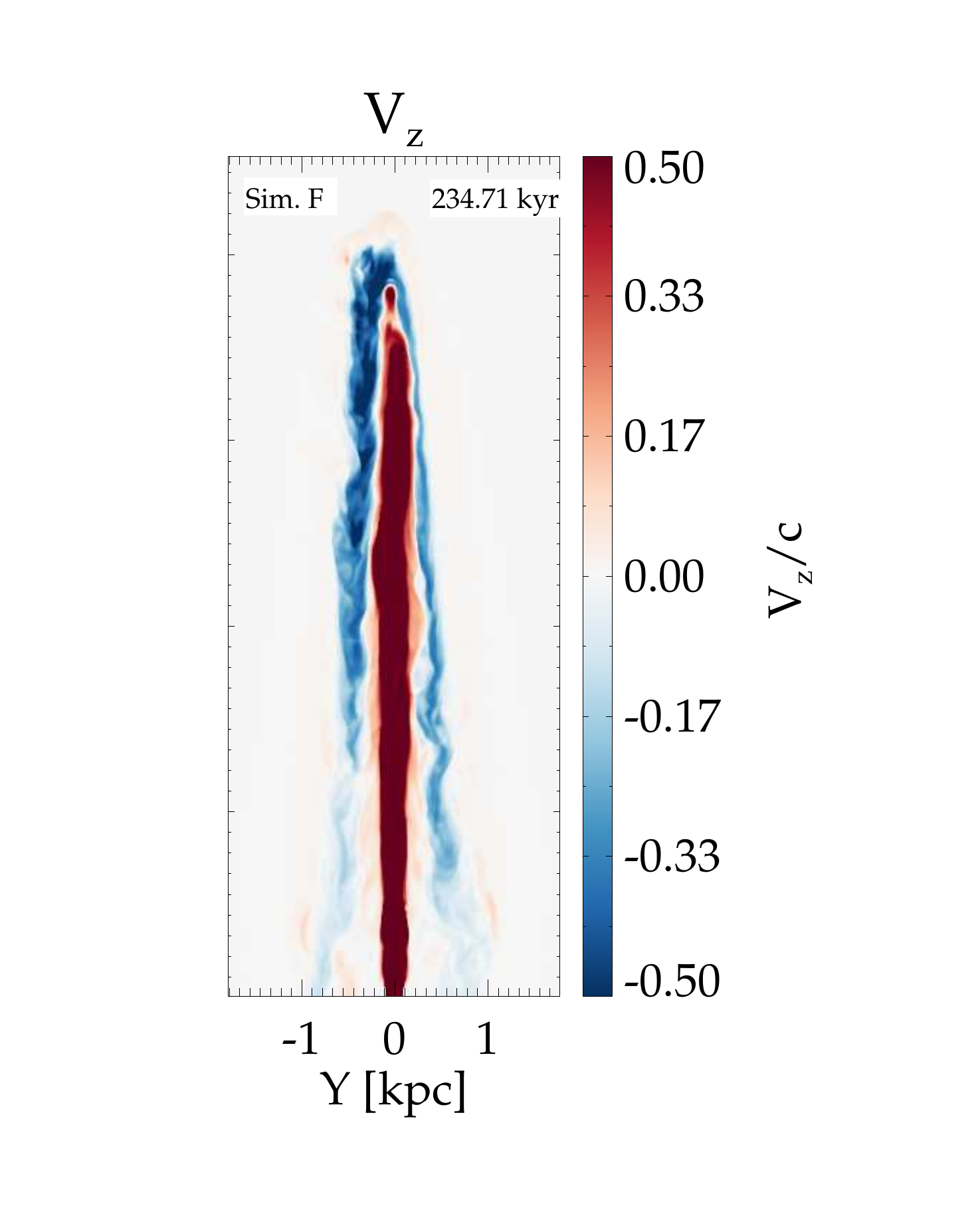}
        \caption{\small The $v_z$ component of the velocity field (normalised to $c$) for simulations D and F. Simulation F shows an extended backflow that remains mildly relativistic ($\sim 0.3 - 0.5$c) for a major part of the cocoon. For simulation D, the backflow loses momentum after a few kpc from the current location of the jet-head.}
	\label{fig.backflow}
\end{figure}
The cocoon has a wide distribution of CRE ages due to their different propagation history. This is accentuated in a turbulent jet where regular streamlines inside the backflow are disrupted due to instabilities. In Fig.~\ref{fig.2DpdfTage} we present the 2D distribution of $\gamma_{\maxrm}$ and the CRE age since injection into the computational domain, normalised by the simulation run-time. The PDFs have been constructed by extracting all particles at a height of $Z = 5\pm 0.2$ kpc, which includes both particles within the jet-spine and the cocoon. The above has been performed for simulations D and F,  at the end of the respective simulations, when the jet has reached a height of $\sim 9$ kpc, such that the height chosen to extract the CREs approximately represents the middle of the jet-cocoon structure.

We firstly notice a bi-modal distribution, with the lower values in the green box representing CREs inside the jet. These CREs have been recently injected, and hence the low age. They travel at near light-speed up to a height of $\sim 5$ kpc, as can be seen from the green horizontal lines in the figures, showing $t_{\rm age} =  5\mbox{ kpc}/(c T_{\rm sim})$, where $T_{\rm sim}$ is the end time of the simulation. The above line represents the lower-end of the distribution, with the rest of the particles being older owing to slower propagation speed inside the jet-axis.

Above the CREs in the jet beam, there is an extended distribution of CREs, which represent the CREs inside the cocoon.The cocoon CREs at $\sim 5$ kpc cover wide range in ages, demarcated by the two limits, $t_L$ and $t_U$ respectively, as denoted in the figure. The upper limit, $t_U$, from the oldest CREs at $Z=5$ kpc, correspond to particles that have exited the jet spine when the jet-head had reached a height of $\sim 5$ kpc during its evolution, and have remained at the similar height up to the end of the simulation. In the regular back-flow model, one would expect the CREs to then stream downwards into the cocoon. However, several CREs follow complex trajectories, which are often lateral, as also shown in Fig.~\ref{fig.pgammaSimB} and Fig.~\ref{fig.simD}. These CREs hover around the original height from which they had exited the jet, which likely result from internal turbulent flows. Being older, they also have a lower value of $\gamma_{\maxrm} \lesssim 10^7$, due to radiative losses. The lower limit, $t_L$, correspond to CREs that have exited the jet at a more recent time, when the jet has evolved to a larger height, and subsequently travelled downwards to $Z\sim5$ kpc with the backflow.

The above age ranges can be better understood by constructing an approximate model of the time spent by a CRE from its injection to its final position in a backflow. Suppose the CRE after injection at the base of the jet, travels along along the flow with a mean propagation speed $v_j$, and exits the main jet-flow at a height $Z^*$ at a time $t^*$ after encountering the jet-head. The CRE then must have been injected into the jet-stream at an earlier time $ t_{\rm inj} =  t^* - (Z^*/v_j)$, where $Z^*/v_j$ denotes the travel time within the jet-axis. Then if the end of the simulation (or in other words the current age of the jet) is $T_{\rm sim}$, the age of the CRE will be:
\begin{equation}\label{eq.tage1}
t_{\rm age} = T_{\rm sim} - t_{\rm inj} =  T_{\rm sim} - \left(t^* - (Z^*/v_j)\right) . 
\end{equation}
After exiting the jet, the CRE freely streams down along a regular backflow. If the mean backflow velocity is $v_b$, then for a CRE at a height $L$ in the backflow (measured from the central source): 
\begin{equation}\label{eq.tage2}
v_b(T-t^*) = Z^* - L
\end{equation}
If we assume the mean propagation speed of the jet-head to be $v_h$, such that $t^* = Z^*/v_h$, then combining \eq{eq.tage1} and \eq{eq.tage2} to eliminate $Z^*$, we get
\begin{equation}\label{eq.tage3}
t_{\rm age} = T - \left(T + \frac{L}{v_b}\right)\frac{(v_j-v_h)}{(v_b + v_h)} \left(\frac{v_b}{v_j}\right)
\end{equation}
\begin{table}
\caption{Estimates of $t_{U}$ from \eq{eq.tage1}}\label{tab.tU}
\centering
\begin{tabular}{| l | c | c | c | c |}
\hline
Sim. 	       &$t_*$  & $T_{\rm sim}$  & $v_j/c$     &  $t_{\rm age}/T_{\rm sim}$  \\
label	       & (kyr) & (kyr)          &             &       \\ 
\hline
D              & 205   & 391            & 0.35        & 0.6    \\
F              & 166   & 235            & 0.7         & 0.39   \\
\hline
\end{tabular} 
\end{table}
Eq~(\ref{eq.tage1})--(\ref{eq.tage3}) together give an approximate estimate of the time taken by the CRE to reach a certain height in the backflow. For CREs at $t_U$, $Z_* = 5$ kpc, and $t_*$ is the time at which the jet reached a height of $\sim 5$ kpc. Using approximate estimates of mean advance speed within the jet $v_j$, one gets approximate estimates of $t_{\rm age}$ which well match with Fig.~\ref{fig.2DpdfTage}, as shown in Table~\ref{tab.tU}.  Note a lower value of propagation speed of the CRE is used for simulation D. This results from the deceleration of the jet-head due to Kelvin-Helmholtz instabilities, resulting in a slower mean propagation speed of the CRE inside the jet axis.

Similarly, to model the ages of the CREs at $t_L$, assuming $(v_j,v_b,v_h) \equiv (0.9,0.3,0.065)c$ for simulation D, gives $t_{\rm age} \sim 0.13 T_{\rm sim}$, for the chosen height of $L = 5$ kpc. This is close to the observed limit in Fig.~\ref{fig.2DpdfTage}. Similarly, for simulation F, the choice of  $(v_j,v_b,v_h) \equiv (0.9,0.3,0.1)c$ gives $t_{\rm age} \sim 0.18 T_{\rm sim}$. The above choices though adhoc, are reasonable in their values and are indicative of the nature of the CRE motion in the backflow. In Fig.~\ref{fig.backflow} we show the $Z$ component of the velocity, depicting the velocity in the jet as well as the backflow. It can be seen that the choice of $v_j \sim 0.9c$ and $v_b \sim 0.3c$ are within limits of the actual values inferred from the velocity maps in Fig.~\ref{fig.backflow}. A smaller value of $v_h$ is used for simulation D to account for the decelerated jet advance due to Kelvin-Helmholtz instabilities. The value used is consistent with the mean advance speed of the jet presented in Fig~15 of \citet{mukherjee20a}.

An interesting point to note is the nature of the distribution of maximum heights reached by CREs with $t_{\rm age} = t_L$ for simulations D and F, as shown in the right panel of Fig.~\ref{fig.2DpdfTage}. While simulation F has a sharp peak at $Z_{\maxrm} \sim 7$ kpc, similar CREs in simulation D has a broader distribution, ranging between $\sim 6 - 8$ kpc. This points to the fact that simulation F being stable to MHD instabilities, has a regular, well-defined backflow, as can also be seen in Fig.~\ref{fig.backflow}. However, MHD instabilities in simulation D disrupt the backflow resulting in intermittent flow patterns and mean speed for different CRE stream-lines. This results in a wider distribution of CREs of different ages at a given height.

\subsection{Evolution of CRE spectrum}
\subsubsection{Spectrum as a function of time for a single CRE}\label{sec.specsingle}
\begin{figure}
	\centering
	\includegraphics[width = 8 cm, keepaspectratio] 
	{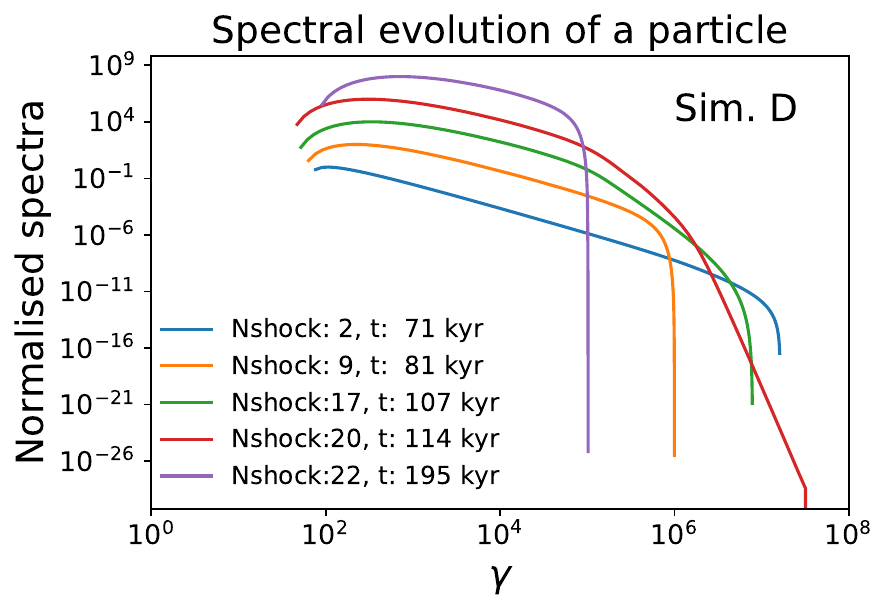}
       \caption{\small The evolution of the spectra for particle P1 in simulation D from Fig.~\ref{fig.simD}. The legends show the number of shocks crossed and the simulation time for each spectrum. The spectra are normalised to their maximum value. Each spectrum is offset vertically by a factor of 100 from the one below for better visual representation.}
	\label{fig.specevol}
\end{figure}
The spectrum of a CRE macro-particle changes in the simulation due to two reasons: i) shock encounters that accelerate the electrons and ii) radiative losses due to synchrotron or inverse-Compton emission. Losses due to inverse-Compton interaction with CMB are nearly constant at all locations, and secondary to synchrotron driven cooling for strong magnetic fields  \citep[$B \gtrsim B_{\rm CMB}$, as in][]{ghisellini14a}. For the simulations explored in this work, the magnetic field is well above the critical field. Fig.~\ref{fig.specevol} shows the evolution of the spectrum of particle P1 of simulations D whose trajectory has been shown in Fig.~\ref{fig.simD}. The figures show how the spectrum changes after the particle has experienced multiple shock encounters. 

At the initial stages the spectra are well represented by a power-law with a sharp cut off due to cooling losses, as expected from a cooling population of shocked electrons \citep{harwood13a}. However, the energy distributions at the intermediate times are not described by a simple power-law with an exponential cut-off. Some of the spectra have a curved shape, well approximated by piece-wise power-laws (for example the red and green curves). Such an evolution may arise when multiple shocks of varying strengths are encountered, as was demonstrated in the top panel of Fig.~\ref{fig.mockdsa},  in Sec.~\ref{sec.convol}. However, the high energy tail will eventually decay with time and the softer slopes from the strongest shock encountered during the particle's history will set the slope at lower energies. Eventually the spectra at end (purple curve) can be approximated as a power-law with an exponential cut-off. 

\subsubsection{Average CRE spectrum at different locations}\label{sec.averageSpectra}
\begin{figure}
	\centering
	\includegraphics[width = 8 cm, keepaspectratio] 
	{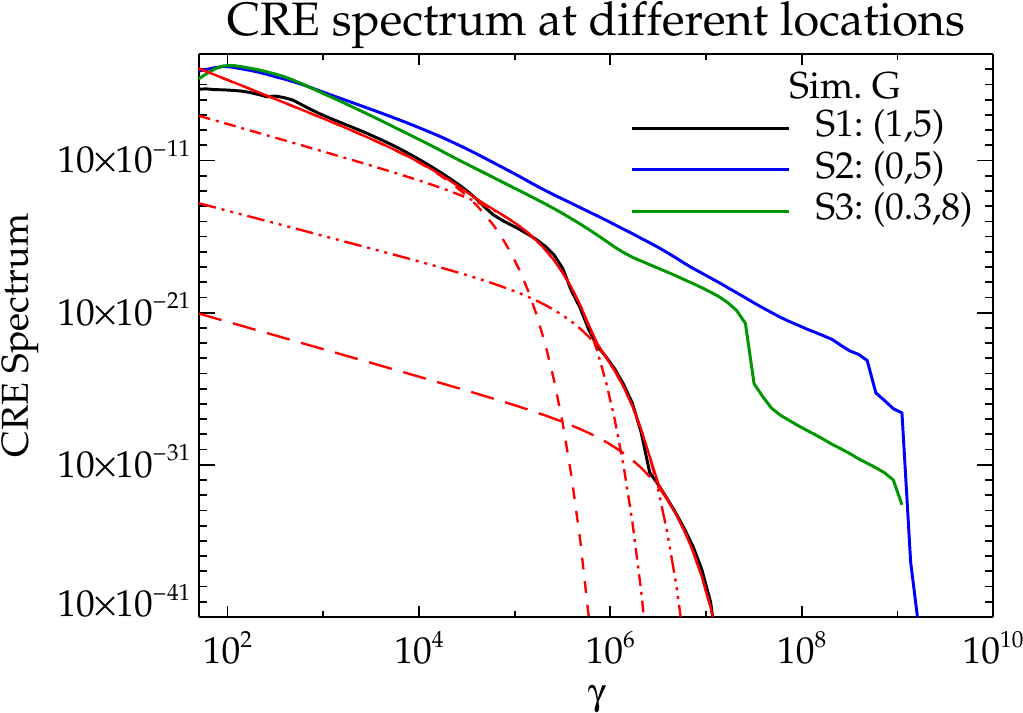}
        \caption{\small The total CRE spectra of particles at three different locations for simulation G at the end of its runtime. The coordinates (in kpc) of the centre of the chosen locations in the ($Y-Z$) plane are presented in the figure legend. The black and blue curves are centred at the cocoon and the jet respectively. The green curve is near the jet head, slightly off-centred from the $Z$ axis. The three red curves in dashes and dots are representative power-laws with exponential cut-offs, with the parameters in Table~\ref{tab.expcutoffparams}, whose envelope given by the red solid line well approximates the spectrum of P1 (in black). See text in Sec.~\ref{sec.averageSpectra} for further details.}
	\label{fig.spectraG}
\end{figure}
CREs have significant difference in spectra depending upon their shock crossing histories during their trajectory due to internal turbulence in the cocoon. The resulting spectra of a collection of non-thermal electrons over a region can thus strongly deviate from a power-law with a sharp cut-off. We demonstrate an example of this in Fig.~\ref{fig.spectraG}, where we show the total particle spectra centred at three locations in the $Y-Z$ plane. Particles in a volume consisting of an area of $200\times200 \mbox{ pc}^2$ in the $Y-Z$ plane and the entire $X$ axis are then extracted and their spectra summed and plotted in Fig.~\ref{fig.spectraG}. The first two locations (S1, S2) represent the spectra expected from the cocoon and the jet axis respectively at $Z \sim 5$ kpc. The third location (S3) samples the spectra from a region near the jet-head.

\begin{table}\label{tab.expcutoffparams}
\centering
\caption{Parameters of spectral component in Fig.~\ref{fig.spectraG}, given by a power-law with exponential cut-off described in \eq{eq.mockdsa}. }\label{tab.expcutoffparams}
\begin{tabular}{| l | c | c |}
\hline
Component  &  Slope ($\alpha$)  &  Cut-off Lorentz factor ($\gamma_c$)   \\     
\hline                                                                                      
1          &  2.5               &  $10^4$   \\
2          &  1.8               &  $10^{4.6}$ \\
3          &  1.65              &  $10^{5.1}$ \\
4          &  1.8               &  $10^{5.7}$ \\
\hline
\end{tabular} 
\flushleft
\end{table}
The locations S2 and S3 being near the jet axis, have their spectra  extended to $\gamma \sim 10^9$, as the CREs are accelerated inside the jet. On the other hand, the location S1, sampling the cocoon, has a decaying spectrum (in black). The decay in the spectrum is not a sharp exponential cut-off as often expected for cooling electrons \citep{harwood13a,harwood15a}, and also seen for single electrons in Fig.~\ref{fig.specevol}. Instead, one can identify clear bends in the spectra which are indicative of a super-position of different electron populations, with different shock and cooling histories. We over-plot in red four representative power-law curves with exponential cut-offs following \eq{eq.mockdsa}. The parameters are listed in Table~\ref{tab.expcutoffparams}. The curves have different cut-off Lorentz factors indicating difference in cooling time scales, and slightly different power-law index implying different shock crossing histories. The superposition of these individual components, which we reiterate is not an actual fit, represents well the final spectrum. 

This indicates that the spectrum of a region in the cocoon will have imprints of different electron populations, which arise out of different evolutionary histories of the CREs in a turbulent cocoon. A similar such imprint of multiple electron population is seen in for S3 at $Z \sim 8$ kpc of Fig.~\ref{fig.spectraG}. It clearly has two components, with a slightly older population having a cut-off Lorentz factor at $\gamma \sim 10^{7.5}$, and another population of shocked electrons extending as a power-law up to $\gamma \sim 10^9$. 
Since the location of S2 is close to the jet axis, the sum of the spectra of all CREs along the $X$ axis will consist of CREs both inside the jet-axis, as well as the intervening cocoon with backflow. The resultant spectrum gives a complex shape with two distinct populations, one slightly cooled in the cocoon and the other from the jet spine.
Similarly, also at location S2, we observe the presence of two components, with cut-offs that are, however, at quite close values of the  Lorentz factor.

\subsection{Particle energy distribution}\label{sec.equipart}
\begin{figure*}
	\centering
	\includegraphics[width = 12 cm, keepaspectratio] 
	{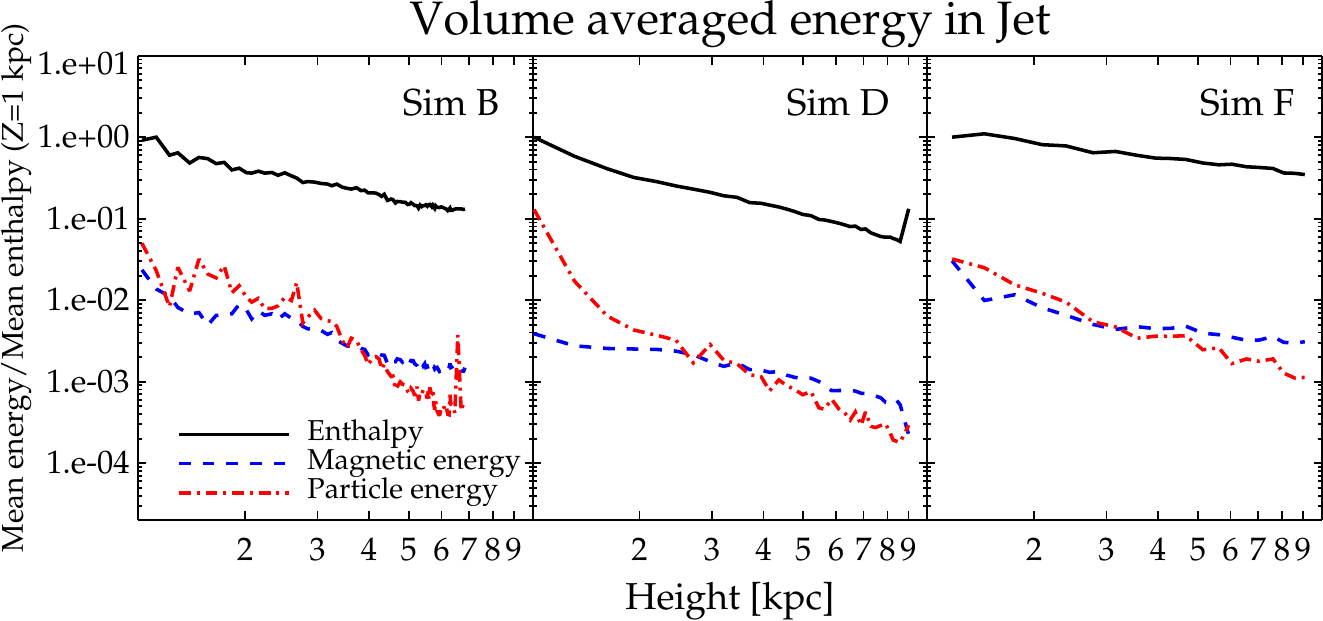}
	\includegraphics[width = 12 cm, keepaspectratio] 
	{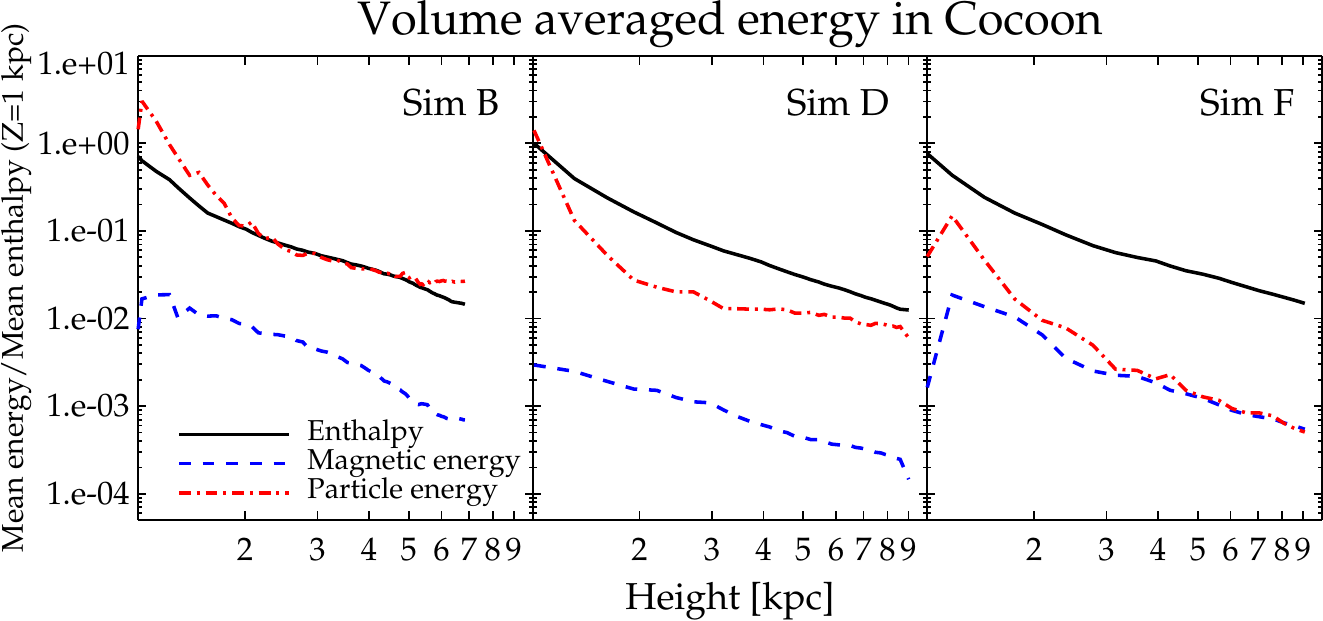}
        \caption{\small The evolution of the volume averaged enthalpy (black), magnetic energy (blue dashed) and particle energy (red dash-dotted) with jet height for three simulations. \textbf{Top:} The values computed for the jet ($\mbox{Jet Tracer } \gtrsim 0.8$). \textbf{Bottom:} The energy components for the cocoon ($10^{-7} \lesssim \mbox{Jet Tracer} \lesssim 0.8$). The curves have been normalised to the starting value corresponding to the jet height $Z \sim 1$ kpc.  }
	\label{fig.engcompare}
\end{figure*}
An important ingredient in modelling the evolution and emission from lobes and jets of radio galaxies is the inherent assumption that the energies in fluid enthalpy, magnetic field and non-thermal particles are in approximate balance, the so-called equipartition argument \citep{hardcastle02a,croston05a,worrall06a,ineson17a}. Although a convenient assumption in most cases, it is unclear whether such a balance is actually reached and to what extent jet parameters and fluid instabilities can affect it. We present in Fig.~\ref{fig.engcompare} the evolution of the volume averaged values of the three different energy components in the jet and cocoon separately. The relativistic fluid enthalpy is given by $\rho h  = \rho c^2 + \rho \epsilon + p$, where the internal energy ($\rho \epsilon$) for a Taub-Matthews equation of state is given by \eq{eq.TM}. The values are computed from a time when the jet height is larger than $Z_j \gtrsim 1$ kpc. The regions with jet tracer between $10^{-7} \lesssim \mbox{ Tracer} \lesssim 0.8$ is considered to be the cocoon, and the volume with $\mbox{Tracer } \gtrsim 0.8$ is considered to be the jet.

\begin{enumerate}
\item{\textbf{Jet beam:}} 
We first note that inside the jet, the fluid enthalpy dominates over both components. The magnetic field and the fluid enthalpy maintain a steady  difference for all simulations. The particle energy is comparable to the magnetic field energy, but lower than the fluid enthalpy by an order of magnitude or more. This is by design, and results from the choice of $f_E = 0.1$ to be enforced at the shocks, as described by \eq{eq.fracE} in Sec.~\ref{sec.convol}. However, simulation D, shows some noticeable difference. It is to be noted that simulation D has a lower magnetisation ($\sigma_B = 0.01$) and hence a lower magnetic energy density to start with. Thus the CRE energy is higher than the magnetic field initially, although close to $\sim 0.1$ of the enthalpy. However, over time, the particle energy decreases and becomes comparable to the magnetic field energy inside the jet.
\item{\textbf{Cocoon region:}}
\begin{itemize}
\item{\emph{Stable jets:}}
In simulation F (right panel of Fig.~\ref{fig.engcompare}), the particle energy is again lower than the fluid enthalpy and maintains a steady difference, as in the jet. This is because the particles are shocked at the jet head and stream down with the backflow and cool in the cocoon. The particle energy density is further diluted by the expansion of the cocoon volume due to the rapid advance of the jet. This is typical of the standard expectations of evolution of an FRII jet \citep{jaffe73a,turner18a}.

\item{\emph{Unstable jets:}}
Inside the cocoon of unstable jets such as the kink unstable simulation B (left panel) and the KH unstable simulation D (middle panel), the particle energy can become comparable to the fluid enthalpy. For simulation B this results from the following two effects: i) the particles are strongly energised by an extended shock due to the bent jet-head at different times; ii) instabilities slow down the jet advance speed and the cocoon is continuously fed with energetic particles, which may be further shocked due to internal turbulence. Thus low power unstable jets with slow advance speed can have increased CRE energy density. 

Similarly, in simulation D (middle panel of Fig.~\ref{fig.engcompare}), the particle energy shows an initial decline followed by an increase to become comparable again to the fluid enthalpy. The initial decline results from the adiabatic expansion of the cocoon as the jet advances. The subsequent rise occurs after the onset of the KH instabilities, when the jet has grown sufficiently. The instabilities both slow down the jet advance and accelerate the CREs at shocks due to the ensuing turbulence, which results in pumping of energetic CREs into the cocoon. This makes the CRE energy density eventually become comparable to the fluid enthalpy itself.
\end{itemize}

It is to be noted that in the jet beam, where the CRE encounter strong shocks, the enthalpy always exceeds the particle energies. This follows from our imposed constraint of the CRE energy being a fraction of the internal energy ($f_E = 0.1$) at shocks. However, in the settling backflows, the cocoon of unstable jets can have an excess build up of high energy CRE due to slow expansion of the cocoon and re-acceleration at internal shocks.
\end{enumerate}

\begin{figure*}
	\centering
	\includegraphics[height = 7.2 cm, keepaspectratio] 
	{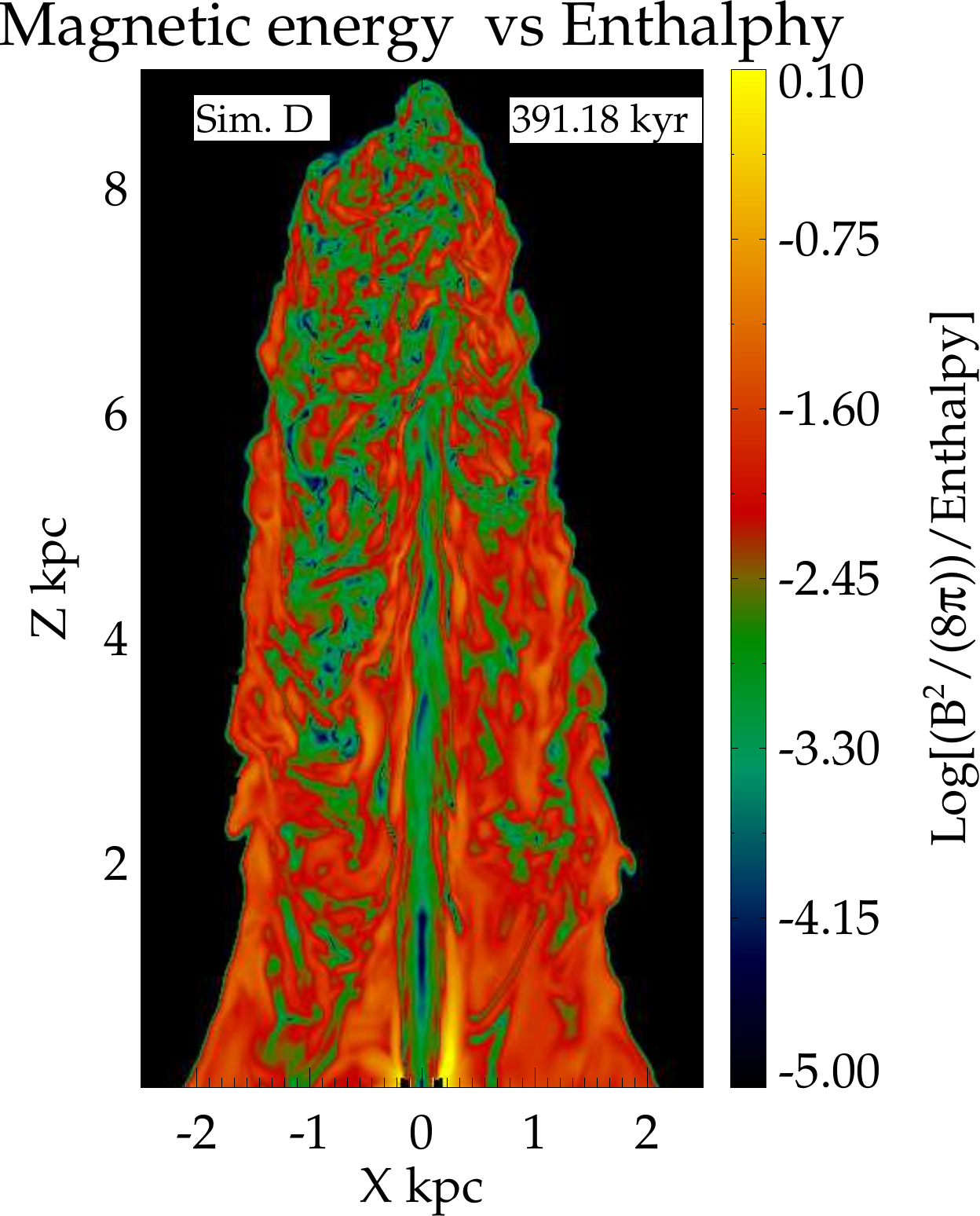}\hspace{-0.cm}
	\includegraphics[height = 7.2 cm, keepaspectratio] 
	{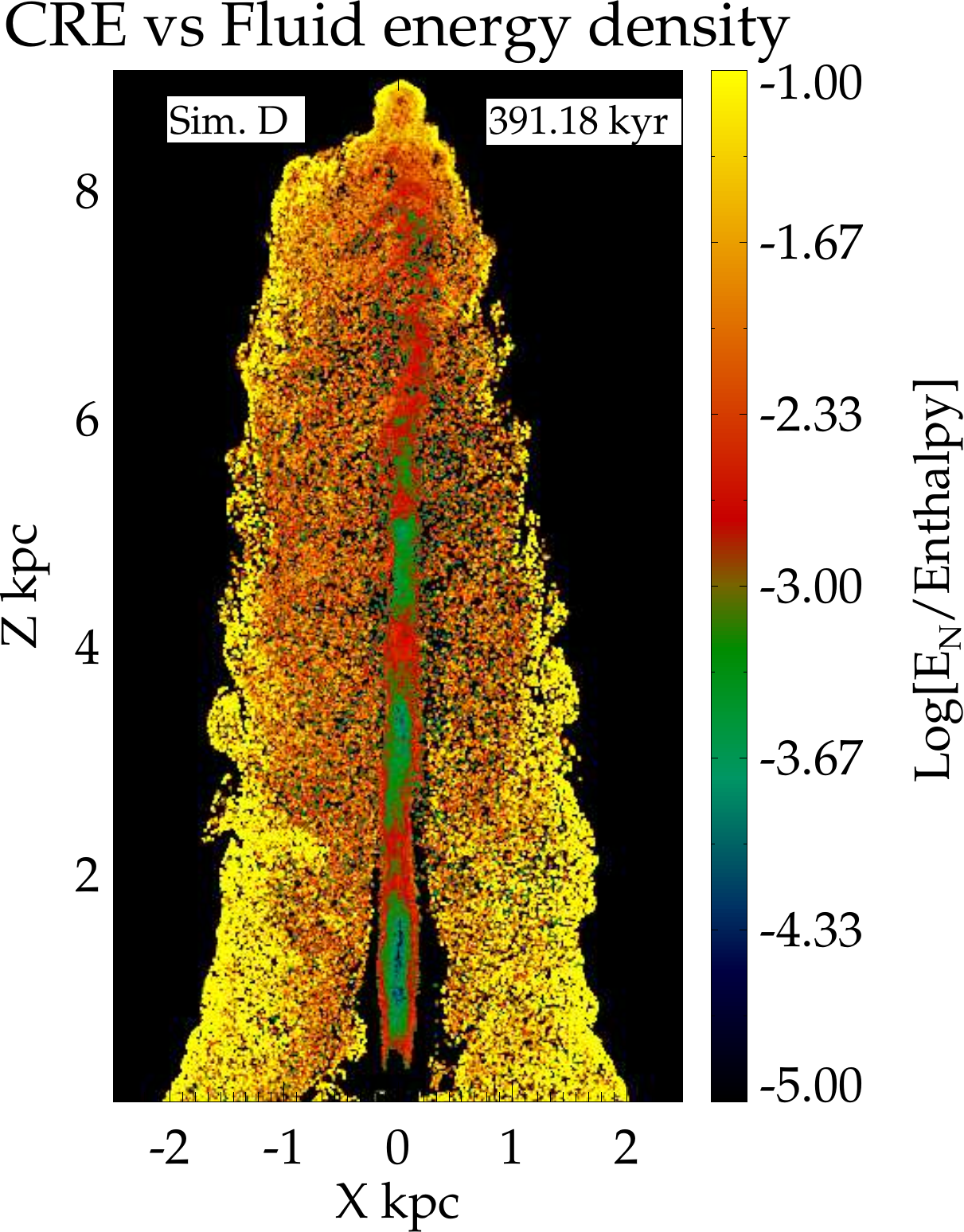}\hspace{-0.cm}
	\includegraphics[height = 7.2 cm, keepaspectratio] 
	{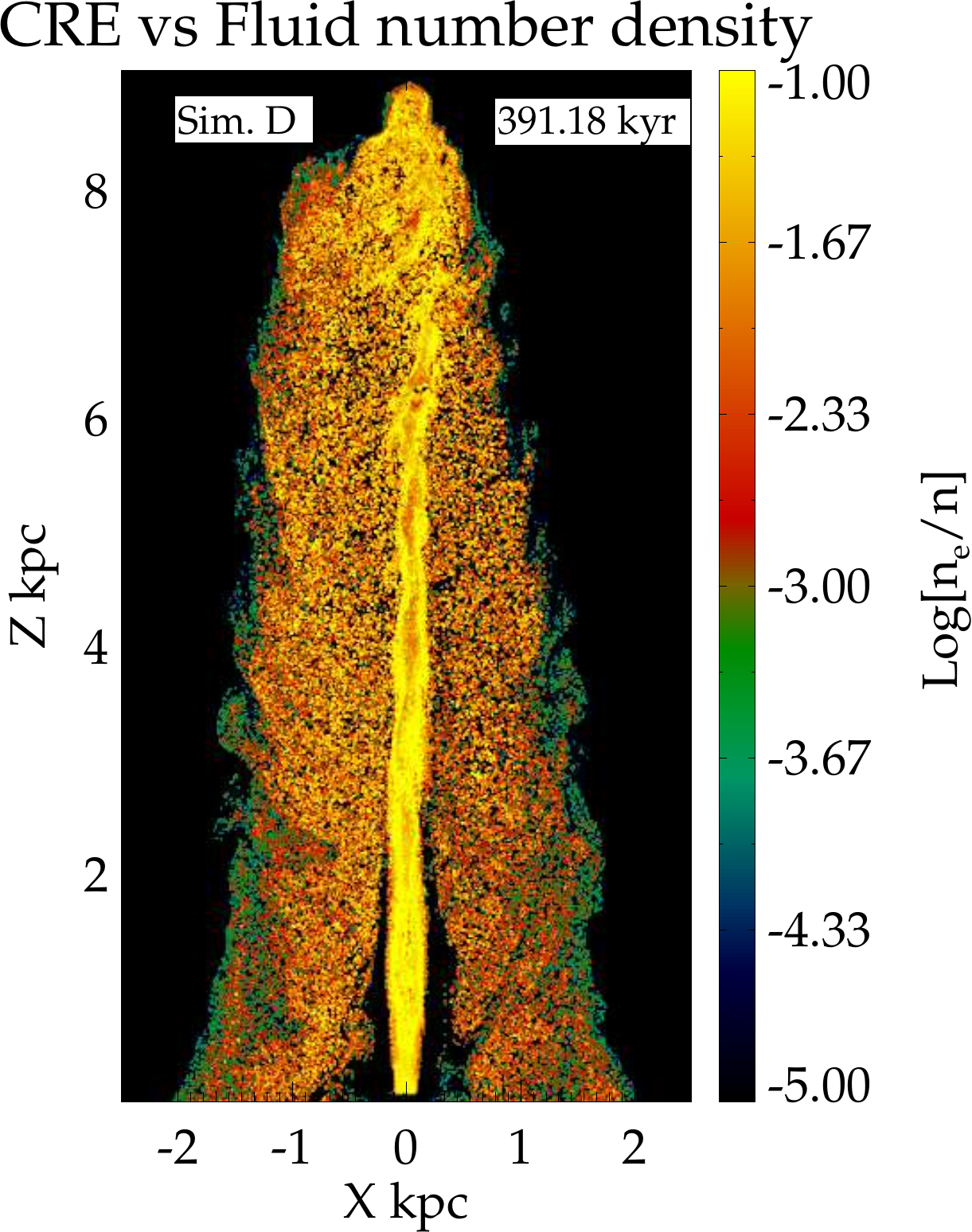}\hspace{-0.cm}
        \caption{\small Figures demonstrating fractional energies between different components as a measure of equipartition. \textbf{Left:} Ratio of the magnetic field energy density to the relativistic fluid enthalpy. \textbf{Middle:} Ratio of the CRE energy density to the fluid enthalpy. \textbf{Right:} Ratio of the CRE number density to the fluid number density. See text in Sec.~\ref{sec.equipart} for details. }
	\label{fig.equipart}
\end{figure*}
The distribution of the different energy components inside the jet is shown more explicitly in Fig.~\ref{fig.equipart}. We present the maps of the ratio of magnetic energy to the fluid enthalpy in the $Y-Z$ plane for simulation D. As also shown by Fig.~\ref{fig.engcompare}, the magnetic field remains secondary to the fluid enthalpy since the magnetisation of the injected jet for simulation D is $\sigma_B = 0.01$. However, the cocoon has local patches of high and low values, indicating small scale local turbulence. This has been covered in extensive detail in Sec.~3.2.2 of Paper I. 

The middle panel of Fig.~\ref{fig.equipart} shows the CRE energy deposited on the computational grid using the ``Cloud-in-Cell" (CIC) approach of weighting, as described in Sec.~3.3 of \citet{mignone18a}. Inside the jet axis, the particle energy is enhanced at certain locations (in red) with lower energy islands following them. These regions of enhanced particle energy correspond to recollimation shocks inside the jet beam where the particles are energised and accelerated. In the cocoon the particle energy has a more homogeneous distribution. The ratio of particle energy to enthalpy is larger at the edges of the cocoon, due to decline of the fluid pressure. The particle number density is highest inside the jet axis, and uniformly distributed inside the cocoon. The ratio of particle number density to the fluid is capped at $0.1$ due to enforced choice of $f_N = 0.1$ in \eq{eq.fracN} in the present simulations.

We have to caution that the results presented in this section have some limitations due to their dependence on the somewhat arbitrary choices of the two parameters $f_E$ and $f_N$. However the general qualitative trends shown in this section are expected to hold. Additionally, we do not account for the back-reaction of the relativistic particle population on the fluid, although in some locations their energy density becomes comparable to the fluid enthalpy. Recent works \citep[e.g.][]{bromberg09a,bodo18a} have shown that radiative losses can affect the structure of flow in shocked regions. However, in our simulations, fluid enthalpy becomes comparable to the CRE energy density only at certain regions in the cocoon of unstable jets, which are not expected to harbour strong shocks. Hence, the lack of radiative losses in our simulations are not expected to strongly affect the flow dynamics and the qualitative results presented here. The effect of coupling radiative losses to the fluid equations however will be addressed in future works.

\section{Discussion and summary}\label{sec.discuss}
In this paper we have presented 3D time-dependent numerical simulations of non-thermal cosmic ray electrons (CREs) embedded in an ideal relativistic magnetohydrodynamic flow jets of typical length scales $\sim 10$ kpc. By exploring different jet conditions, we have focused on the impact of MHD instabilities not only on the overall jet morphological structure but also on the spatial distribution of the CREs along the jet beam and its backflow, and the corresponding impact on the CRE spectrum.

Our results aim at surpassing semi-analytical models, previously proposed to describe the evolution of the CRE spectrum in the cocoon backflow \citep[viz. the KP, JP and Tribble models, see][for a review]{harwood13a}.  Indeed, the vast majority of these  models rely on several simplistic assumptions for ease of computation, such as constant magnetic field in the cocoon and constant pitch angle \citep[KP model, ][]{pacholczyk70a},  or  a constant magnetic field and isotropic pitch angle \citep[JP model ][]{jaffe73a},  or a magnetic field distribution evolving with time due to adiabatic expansion of the cocoon \citep{murgia99a} or a turbulent magnetic field distribution in the cocoon \citep{tribble91a,hardcastle13a}. 
However, a time-dependent scenario, such as the one explored here, is likely to create complex flow features where non-thermal electrons can experience multiple shock crossings with non-trivial geometries. Last but not least, the effect of fluid instability can considerably affect the magnetic field strength in localised regions of space. 

We can summarise our primary results and their implications as follows.
\begin{enumerate}
\item \emph{Stable jets:}  Non-thermal electrons in stable jets show the expected trends (as in Fig.~\ref{fig.jetCartoon}) of being first accelerated at recollimation shocks inside the jet axis and then the jet-head, followed by settling motions in the backflow of the cocoon. This is similar to the classic CRE evolutionary models proposed in early works \citep{jaffe73a} where the jet axis and hotspot are filled with young energised electrons and the cocoon has older cooler electrons. 

\item \emph{Multiple shock encounter in unstable jets:} CREs undergo repeated  acceleration in unstable jets at shocks of varying strengths and magnetic field. This can leave an imprint on the slope and maximum energy of the spectra. Kink instabilities lead to bending of the jet-head which cause complex shock structures with extended curved features. The CREs thus experience repeated acceleration events before entering the backflow.  Kelvin-Helmholtz (KH) instabilities in low magnetic field jets produces internal turbulence and vortices. This causes the CREs to have complex motions while travelling along the main jet flow as they  spiral in local vortices. Re-accelerations of electrons in turbulent cocoons of unstable jets, coupled with slower expansion speed, also result in excess build of CRE energy in the jet cocoons, as discussed in Sec.~\ref{sec.equipart}. 

Several aspects of the above results, such as the complex shocks in the lobes and multiple shock crossings of particles in the backflow, are complimentary to recent findings from relativistic hydrodynamic simulations of large scale jets by \citet{matthews19a}. Such shocks have been proposed to be conducive to the production of Ultra High Energy Cosmic Rays (UHCER), as shown in \citet{bell19a}. Qualitative agreements between the nature of evolution of the shock structures between our results, though on different physical scales, lends further support to the possibility of accelerating cosmic rays in shocks of extragalactic jets.

Turbulent jets thus create a distinct population of freshly accelerated  electrons which have undergone several encounters with internal shocks. Such CREs are markedly absent in stable jets, indicating that creation of a separate energetic population of electrons is inherently related to the local dynamics of the fluid and the resultant shock acceleration. Presence of a secondary population have been often invoked to explain the excess observed emission at high energy bands of some jets, modelled as synchrotron emission from an energised electron population with $\gamma_e \gtrsim 10^7$ \citep{atoyan04a,kataoka05a,hardcastle06a,jester06a,meyer15a,migliori20a}. Such a scenario has also been proposed by \citet{borse21a} in local simulations of particle accelerations of a section of a non-relativistic jet. In this work we demonstrate that re-acceleration at internal shocks in the cocoon and the jet can indeed create a distinct secondary population of energised electrons as theorised in the above works, and that this is more pronounced in jets prone to MHD instabilities.

An important implication of CREs being energised in the cocoon is that the mean spectral index of the electrons and hence also the synchrotron spectrum, is expected to become shallower. This is similar to the  ``continuous injection" model used to describe spectrum of observed sources \citep[][]{murgia99a,brienza18a} with shallow radio spectral slopes, indicative of fresh injection of accelerated electrons. This will be further  addressed in a forthcoming work, where we will present the observable synchrotron emission expected from our simulations (Paper III, Mukherjee et al. in prep).

\item \emph{Mixing of CRE population:} In the same simulation, CREs with different trajectories can experience very different fluid conditions, which affect their spectrum, as shown for particles P2 and P3 of simulation B in Sec.~\ref{sec.gmmhist}. CREs can be accelerated to high energies both if they cross strong shocks or if they pass through regions of low magnetic field where slower synchrotron cooling time scales result in more efficient acceleration. Thus at a given height, there will be mixing of different population of electrons, with  different evolutionary history.  The distribution of CRE ages in the cocoon since their injection in the simulation, is well described by approximate analytical model of backflow presented in Sec.~\ref{sec.tage}.

\item \emph{Impact of shock encounters on CRE spectrum:} The spectrum of a CRE that has undergone multiple shock crossings is a piece-wise power-law (see Sec.~\ref{sec.specsingle}). The cooling spectrum, however, finally attains the shape of a power-law with exponential cut-off, with the slope at lower energies being determined by its strongest shock encounter. Since CREs with different shock histories lie at similar locations, spatially averaged spectrum of a region will have imprints from different cooling populations. This can cause the spectrum to have complex shapes that are best modelled by superposition of different cut-off power-laws due to the different CREs (e.g. Fig.~\ref{fig.spectraG}). The resultant envelope shows a curved spectrum at the higher energies. Such results have also been hinted in earlier works such \citet{micono99a} and \citet{meli13a}, which have modelled multiple shock encounters of electrons at different recollimation shocks inside the jet axis. Our results show that this will be expected also in the cocoons of turbulent jets. Observed curvature in radio spectrum of some sources have been attributed to intrinsic property of the electron distribution itself by some recent papers, such as \citet{duffy12a,nyland20a}. Such curvature at higher energies of the spectrum naturally arises in our simulations due to multiple shocked populations. 

\end{enumerate}

In the current work, we have focused on presenting the results of the dynamics and spectral evolution of non-thermal electrons injected with the jet and the impact of MHD instabilities. In future papers, we will present the implications of these results on the observable non-thermal emission arising from processes such as Synchrotron and Inverse-Compton; and how differences in jet-dynamics and internal fluid properties affect such emission processes.

\section{Acknowledgement}
We thank the referee for their constructive comments which helped to improve the clarity of the paper. We acknowledge support by CINECA through the Italian Super Computing Resource Allocation (ISCRA) scheme and by the Accordo Quadro INAF-CINECA 2017 for the availability of high- performance computing resources. The authors acknowledge support from the PRIN-MIUR project Multi-scale Simulations of High- Energy Astrophysical Plasmas (Prot. 2015L5EE2Y). BV would like to acknowledge the support provided from Max Planck Society (MPG) in establishing Max Planck Partner Group at IIT Indore.

\section{Data Availability}
The derived data generated in this research will be shared on reasonable request to the corresponding author.

\appendix
\section{A summary of the method to evolve CREs in simulations}\label{append.DSA}
\subsection{Phase space evolution of a CRE not in a shock}\label{append.bv18}
The CREs are evolved both in space and energy following the 6 dimensional Boltzmann transport equation \citep{webb89a,mimica09a,vaidya18a} after ignoring the effect of spatial diffusion, shear and viscous energy exchanges, Fermi second order processes and non-inertial energy changes as outlined in section 2 of \citet[][hereafter BV18]{vaidya18a}. The number of particles per unit volume in the energy range $(E,E+dE)$ is given by $N(E,t)dE$. Each CRE macro-particle represents an ensemble of non-thermal electrons with an energy distribution defined by a spectrum, which in our work is discretised into finite logarithmically placed energy bins. For this work, the spectra have been defined on 100 bins. The spatial location ($x_p$) of a macro-particle is updated by solving $dx_p/dt = v(x_p)$, where $v(x_p)$ is the velocity of the fluid interpolated onto the location of the CRE macro-particle.

As outlined in section 2 of BV18, the Boltzmann transport equation can be decomposed into characteristics, which can be solved to show that the particle number density $N_{p0}(E_0,\tau$) between an energy interval $dE_0$ evolves as 
\begin{equation}
N_p(E,\tau) dE = N_p(E_0,0)dE_0. \label{eq.chi} 
\end{equation}
The energy along the characteristic follows:
\begin{equation}
E(\tau) = \frac{E_0 \exp(-a(\tau))}{1 + b(\tau) E_0} \label{eq.eng}
\end{equation}
Here the $a(\tau)$ is the adiabatic advection coefficient and $b(\tau)$ represents the radiative losses, defined as:
\begin{align}
a(\tau) &= \int_0^\tau \frac{1}{3} (\nabla_\mu u^\mu) d\tau^\prime \\
b(\tau) &= \int_0^\tau \frac{4 \sigma_T c \beta^2}{3 m_e^2 c^4} \left\lbrack \frac{B^2}{8 \pi} + a_{\rm rad} T_0^4(1+z)^4 \right\rbrack e^{-a(\tau^\prime)} d\tau^\prime  
\end{align}

$\tau$ is the proper time which is related to the time in observer frame as $d\tau = dt/\gamma$, $ \gamma$ being the bulk Lorentz factor of the particle. $\sigma_T$ is the Thompson scattering cross section, $\beta$ is the velocity of the particle normalised to the speed of light ($c$), $m_e$ is electron mass, $B$ is the magnetic field interpolated at the location of the particle, $a_{\rm rad}$ is the radiation constant and $T_0 = 2.728$K is the CMB temperature at $z=0$.

Note that \eq{eq.chi} implies that the number density of CRE in the energy bin remains conserved. However, the edges of the energy bins evolve according to \eq{eq.eng}. Thus coupled together, \eq{eq.chi} and \eq{eq.eng} describe the evolution of CREs with an averaged number density of $N_p(E_0,0)$ in an energy bin of range $(E_0,E_0+ dE_0)$ at $\tau=0$, as they evolve to time $\tau$. In practise, the spectrum of a CRE is updated at each time step by evolving the energy bins of the discretised spectrum following \eq{eq.eng} while preserving the value of $N_p(E_0,0)$ for the energy bin \citep{vaidya18a,huber21a}.

\subsection{Spectral update at shocks}
\subsubsection{Identifying entry and exit of shocks by CREs}
A CRE is considered to enter shocked cell if the fluid pressure in the computational volume and its neighbours satisfy:
\begin{align}
&\frac{|\Delta p_x| + |\Delta p_y| + |\Delta p_z| }{p_{\rm min}} > \zeta_F = 3 \label{eq.delp}\\
&\mbox{ where } |\Delta p_x| = |p(x_{i+1},y_i,z_i) - p(x_{i-1},y_i,z_i)| \nonumber.
\end{align}
Here $p_{\rm min}$ is the minimum pressure of cells surrounding the given computational cell which the CRE enters. The above gives an approximate estimate of the maximum pressure difference across computational cells. The fluid pressure is interpolated on to the particle as it enters a shock ($p_{\rm upst}$). The CRE trajectory is followed and the fluid pressure interpolated onto the particle ($p_F$) is recorded. The CRE  is considered to have crossed a shock if 
\begin{equation}
(p_F - p_{\rm upst})/p_{\rm upst} > \zeta_{\rm CRE} = 3. 
\end{equation}

The above criteria, which depends on the history of the pressure change experienced by the particle, coupled with that in \eq{eq.delp} are together used to identify if a CRE has successfully entered and exited a shock, and is due for a shock-update based on implementation of diffusive shock acceleration outlined in the next section. 

\subsubsection{Implementation of DSA}\label{append.dsa}
The spectrum of a CRE particle that has exited a shock is updated following a prescription of diffusive shock acceleration \citep[as described in detail in section 2.4 of ][]{vaidya18a}. The shock strength is quantified in terms of its compression ratio, defined as
\begin{equation}
r = \rho_{2}/\rho_{1} = \mathbf{v_{1}}\cdot \boldsymbol{\hat{n}}_s/(\mathbf{v_{2}}\cdot \boldsymbol{\hat{n}}_s) \label{eq.cmpr}
\end{equation}
where ($v_{1}$,$\rho_{1}$) and ($v_{2}$,$\rho_{2}$) are the upstream and downstream velocities and densities respectively, in the shock rest frame. The spectrum of a shocked CRE is updated by convolving the downstream spectrum of the CRE with a power-law spectrum, having a spectral index defined by the theory of diffusive shock acceleration as outlined in Sec.~\ref{sec.convol}. The maximum energy of the new spectrum is determined by equating the time scale of energy loss due to synchrotron radiation to the shock acceleration time scale. The expression is \citep[as in][]{vaidya18a}:
\begin{align}
\gamma_{\rm max} &= \left(\frac{9c^4 m_e^2}{8\pi B a_{\rm acc} e^3}\right)^{1/2} \label{eq.gammamax}\\
a_{\rm acc} &= \frac{\eta r}{\beta_1^{\prime2}(r-1)}\left\lbrack \cos^2 \theta_{B1} + \frac{\sin^2 \theta_{B1}}{1+\eta^2} \right. \nonumber \\
  &+ \left. \frac{rB_1^{\prime}}{B_2^{\prime}}\left(\cos^2 \theta_{B2} + \frac{\sin^2 \theta_{B2}}{1+\eta^2}\right) \right\rbrack \label{eq.acc}
\end{align}
Here $\eta$ is the ratio of the gyro frequency to the scattering frequency \citep{takamoto15a,vaidya18a}, which is set to a constant value of $\eta \simeq 1/\cos(45^\circ)$ in our simulations\footnote{Depending on the orientation of the downstream magnetic field with respect to the shock normal, a shock is considered quasi-perpendicular if $\cos\theta_{B2} \leq 1/\eta$ \citep{takamoto15a}. In this work, we assume a shock with $\theta_{B2} > 45^\circ$ as perpendicular. This sets the assumed value of $\eta$ for the limit $\theta_{B2} = 45^\circ$.}. The variables ($\theta_{B1}$,$\theta_{B2}$), ($B^\prime_1$, $B^\prime_2$) and ($\beta_1$,$\beta_2$) represent the angle of the magnetic field with respect to the shock normal, magnetic field magnitude and velocity normalised to speed of light respectively, with upstream values labelled by 1 and those downstream by 2. The other terms in \eq{eq.gammamax} are standard constants.

The two most important parameters of \eq{eq.acc} are the strength of the magnetic field ($B$) and the shock compression ratio ($r$), defined in \eq{eq.cmpr}. Thus a CRE can be accelerated to high energies if it passes through a strong shock with high compression ratio $r$ and/or low magnetic field, resulting in increased synchrotron cooling time scales, and thus more efficient acceleration. Thus both at the jet-head where the shock strength is high, and as well as in a turbulent cocoon where weak shocks may lie in regions of relatively lower magnetic field strengths, CREs can be accelerated to high energies.

The index of the power-law spectrum with which the upstream spectrum is convolved, is decided based on whether the shock normal is parallel or perpendicular to the down-stream magnetic field. For parallel shocks ($\theta_{B2} \leq 45^\circ$), the spectral index as per the theory of DSA is \citep{keshet05a,vaidya18a}:
\begin{equation}
q_{||} = \frac{3 r}{r-1} + \left(\frac{1-2r}{r-1}\right) \beta^{\prime 2}_2 \label{eq.qpar}.
\end{equation}
For perpendicular shocks ($\theta_{B2} > 45^\circ$), the spectral index is given by \citep{takamoto15a,vaidya18a}:
\begin{equation}
q_{\perp} = \frac{3 r}{r-1} + \frac{9(r + 1)}{20r(r-1)} \eta^2 \beta^{\prime 2}_1 \label{eq.qperp}
\end{equation}
Note that the first term in \eq{eq.qpar} and \eq{eq.qperp} is the spectral index for a non-relativistic shock \citep{blandford78a,drury83a}.

\subsection{Interpolating a CRE spectrum}\label{sec.interpolate}
Spectrum of a CRE is often needed to be interpolated on to a new energy array, such as while performing the integral to convolve the spectra at a shock update described in Sec.~\ref{sec.convol}, or while adding spectra of many different CRE to get a total spectrum from an area, as done in Sec.~\ref{sec.averageSpectra}. While interpolating, first a new logarithmically spaced energy array is created with its end points set appropriately. For example while adding two spectra, the extrema are set to the lowest and the highest of the two spectra respectively. For performing the convolution in Sec.~\ref{sec.convol}, the lowest energy is unchanged, while the highest energy edge defined from the shock acceleration prescription using \eq{eq.gammamax}.

During the interpolation, for each bin of the original spectrum (e.g. $[\epsilon_{il},\epsilon_{ih}]$ for the $i$-th bin), corresponding bin edges [$\epsilon_{il'},\epsilon_{ih'}$] in the interpolated spectrum are identified such that $\epsilon'_{il} \leq \epsilon_{il}$ and $\epsilon_{ih} \leq \epsilon'_{ih}$. The interpolated spectrum is then determined by assuming a flat reconstruction (i.e. $\langle N_i \rangle$ is constant for the $i$-th bin), while ensuring 
\begin{equation}\label{eq.reconstruct}
\int_{\epsilon_{il}}^{\epsilon_{ih}} \langle N_i \rangle d\epsilon = \int_{\epsilon'_{il}}^{\epsilon'_{ih}} \langle N'_i \rangle d\epsilon',
\end{equation}
Here $\langle N_i \rangle$ is the average electron number density in the $i$-th bin of the spectrum and the primed quantities are the corresponding spectral parameters of the new spectrum where the CRE spectrum is being interpolated. A more sophisticated  linear order reconstruction with appropriate slope limiters has been recently implemented in \citet{huber21a}. However, we opt for a simpler method with first order accuracy to save computation time.

\def\apj{ApJ}%
\def\mnras{MNRAS}%
\def\aap{A\&A}%
\def\apjl{ApJ}
\def\physrep{PhR}
\def\apjs{ApJS}
\def\pasa{PASA}
\def\pasj{PASJ}
\def\nat{Nature}
\def\memsai{MmSAI}
\def\aj{AJ}%
\def\aaps{A\&AS}%
\def\iaucirc{IAU~Circ.}%
\def\sovast{Soviet~Ast.}%
\def\apss{Ap\&SS}

\bibliographystyle{mnras_2020}
\bibliography{dmrefs}

\end{document}